\documentclass{article}
\usepackage{fullpage}

\usepackage{ifthen}
\newcommand{\isShort}{true} 
\newcommand{\shortVer}[1]{\ifthenelse{\equal{\isShort}{true}}{{#1}}{}}
\newcommand{\longVer}[1]{\ifthenelse{\equal{\isShort}{false}}{{#1}}{}}

\usepackage[small,bf]{caption}

\usepackage[hang,flushmargin,stable]{footmisc}
\usepackage{graphicx}
\usepackage[nospace]{cite}
\usepackage{amssymb}
\usepackage{amsmath}
\usepackage{amsfonts}
\usepackage{times}
\usepackage{colortbl}
\usepackage{hhline}
\usepackage{color}
\usepackage{url}
\usepackage{subfigure}
\usepackage{tikz}
\usepackage{bbding}
\usepackage[all]{xy}
\usepackage{array}
\usepackage{multirow}
\usepackage{multicol}
\usepackage{grffile}
\usepackage{paralist}
\usepackage{flushend}

\definecolor{greenish}{RGB}{0,89,89}
\definecolor{darkblue}{RGB}{0,69,109}
\usepackage{etoolbox}
\usepackage[bookmarks=false, colorlinks=true, plainpages = false,  linkcolor = darkblue,   citecolor = darkblue, urlcolor = darkblue, filecolor = blue]{hyperref}

\usepackage{indentfirst}

\ifthenelse{\equal{\isShort}{true}}{}{}

\newcommand{\descr}[1]{\medskip \noindent \textbf{#1}}

\newcommand{\equal}{\hspace*{-0.05cm}=\hspace*{-0.05cm}}
\newcommand{\Cset}{\ensuremath{C}}
\newcommand{\Sset}{\ensuremath{S}}

\title{\bf Controlled Data Sharing for\\Collaborative Predictive Blacklisting\footnote{A preliminary version of this paper appears in the Proceedings of DIMVA 2015. This is the full version.}}

\author{Julien Freudiger$^1$ $\;\;\;$ Emiliano De Cristofaro$^2$ \;\;\; Alex Brito$^1$\\[1ex]
{\normalsize $^1$ PARC (a Xerox Company) $\;\;\;$ $^2$ University College London}}
\date{}

\begin{document}
\pagenumbering{arabic}
\thispagestyle{plain}

\maketitle

\begin{abstract} 
Although sharing data across organizations is often advocated as a promising way to enhance cybersecurity, collaborative initiatives are rarely put into practice owing to confidentiality, trust, and liability challenges. In this paper, we investigate whether collaborative threat mitigation can be realized via a {\em controlled data sharing} approach, whereby organizations make informed decisions as to whether or not, and how much, to share. Using appropriate cryptographic tools, entities can estimate the benefits of collaboration and agree on what to share in a {\em privacy-preserving} way, without having to disclose their datasets. 

We focus on collaborative predictive blacklisting, i.e., forecasting attack sources based on one's logs and those contributed by other organizations. We study the impact of different sharing strategies by experimenting on a real-world dataset of two billion suspicious IP addresses collected from Dshield over two months. We find that controlled data sharing yields up to 105\% accuracy improvement on average, while also reducing the false positive rate.\end{abstract}

\section{Introduction}\label{sec:introduction}
Over the past few years, security practitioners and policy makers
have called for sharing of data related to cyber threats and attacks.
Prior work has shown that organizations are exposed to similar vulnerabilities, targeted by the same malevolent actors, and that collaboration could enhance timeliness and accuracy of threat mitigation~\cite{katti2005collaborating,locasto2005towards,soldo}.
The US government recently initiated efforts to encourage the private sector to share
cybersecurity information to improve US cyber defenses~\cite{whitehouse_last}. 
At the same time, the private sector launched community-based initiatives such as the RedSky Alliance~\cite{redsky}, ThreatExchange~\cite{fb}, 
DOMINO~\cite{DOMINO}, and WOMBAT~\cite{wombat}.

However, collaborative security initiatives have had little success
due to the related confidentiality, privacy, trust,
and liability challenges. Sharing security data may 
damage competitivity, reveal negligence, and
expose sensitive and private information.
In fact, data sharing initiatives are often opposed by the privacy community
as potentially harmful to individuals~\cite{guardian},
while organizations have little choice other than establishing
``circles of trust,'' aiming to control potential loss of competitive advantage 
and data exposure. Alas, this creates
the need for lengthy out-of-band processes to establish trust,
which hinders speediness and economic viability of
such initiatives, as highlighted by a recent Federal Communications Commission (FCC) report~\cite{CSRIC}.

\subsection{Problem Statement}
We investigate whether collaborative threat mitigation
can be realized via a {\em controlled data sharing} approach, i.e., 
seeking an effective middle ground between sharing everything
and sharing nothing, and helping organizations make informed decisions about 
{\em whether or not} to share data, and {\em how much}. 

This raises a few compelling research challenges: \vspace{-0.1cm}
\begin{enumerate}
\item How can organizations {\em estimate benefits} of collaboration? What metrics can 
two organizations use to guide the decision as to  whether or not
they should share data? %
\item Can we ensure that benefit estimation occurs in a {\em privacy-preserving way},  
so that organizations do not need to disclose their entire datasets, but only the minimum required amount of information? %
\item Once two organizations decide to collaborate, {\em how much} and {\em what} should they share? %
\end{enumerate}
We address these challenges in the context of
{\em collaborative predictive blacklisting}, whereby different organizations aim to forecast attack sources, based on their firewall and Intrusion Detection Systems (IDS) logs, and also those generated by collaborating organizations.
We model collaboration as a three-step process in which organizations first estimate the benefits of data sharing among each other, then establish partnerships with promising partners, and finally share data with them.
We aim to investigate which collaboration strategies work best, in terms of the resulting improvement in the prediction accuracy and false positive rate.

\subsection{Roadmap}
We experiment with different metrics for estimating the benefits of collaboration, using Jaccard, Pearson, and Cosine  similarity of the logs, as well as the size of their intersection. We also test different degrees of data sharing, e.g., 
sharing everything or only information about attacks entities have in common. One crucial aspect of our 
work is to impose a fundamental constraint: Benefit estimation and
data sharing should occur in a privacy-preserving way,
which we attain via cryptographic tools for secure two-party computation (2PC)~\cite{Yao}. 
As research in 2PC has produced increasingly efficient and practical
implementations, both for general-purpose garbled circuits~\cite{HEKM11} and 
special-purpose protocols (e.g., private set intersection~\cite{PSZ14,DT12,DGT12}), 
the overhead introduced by the privacy protection layer is appreciably low (cf. Section~\ref{subsec:perf}).

Aiming to compare different strategies, we perform an empirical evaluation using a real-world dataset of 2 billion suspicious IP addresses collected from DShield.org~\cite{dshield} over two months. %
This dataset contains a large variety of contributors, which allows us to test the effectiveness of data sharing among diverse groups of victims. We perform a quantitative analysis of this dataset %
in order to identify victims' and attackers' profiles, among other features. This helps us clean the dataset and design a meaningful (controlled) data sharing experiment. %
We repeatedly select $100$ victims at random and measure the
accuracy improvement of the blacklisting algorithm, performing the prediction by means of a standard algorithm based on Exponentially Weighted Moving Average (EWMA)~\cite{soldo}.

Our analysis yields several key findings. We observe that: (1) The more information is available about attackers, the better the prediction, as intuitively expected; (2) Different collaboration strategies yield a large spectrum of prediction accuracy, and in fact, with some strategies, sharing does not help much; (3) Collaborating with other organizations not only helps improve prediction, but also reduces the false positive rate; and (4) sharing information only about common attackers is almost as useful as sharing everything. 
As a results, we conclude that controlled data sharing 
can help organizations find the right balance between indiscriminate sharing and non-collaboration, i.e., sharing just enough data to improve prediction.

\descr{Paper Organization.} The rest of the paper is organized as follows: next Section overviews related work, then, Section~\ref{sec:preliminaries} introduces relevant background information. Section~\ref{sec:framework} presents our controlled data sharing model for collaborative predictive blacklisting, while Section~\ref{sec:dshield} discusses the Dshield dataset used in our experiments. After presenting the results of our experimental analysis in Section~\ref{sec:experiments}, the paper concludes with Section~\ref{sec:conclusion}, and Appendix~\ref{app:dshield} presents additional statistics about the Dshield dataset.

\section{Related Work}\label{sec:related}

Previous work on collaborative intrusion detection has usually employed a centralized system where organizations  contribute data to Trusted Third Parties (TTPs) in return for blacklisting recommendations. 
Zhang et al.~\cite{zhang2008highly} introduce the notion of highly predictive blacklisting for predicting future attacks based on centralized logs, while follow-up work by Soldo et al.~\cite{soldo} improve
by using an implicit recommendation system and further increase  accuracy. 
Although we re-use one of the prediction algorithms from~\cite{soldo}, previous work~\cite{zhang2008highly,soldo} does not take into account privacy and relies on TTPs.

Prior research attempted to mitigate privacy challenges from security data sharing by 
relying on data anonymization and sanitization~\cite{lincoln2004privacy,slagell2005sharing,porras2006large,xu2002prefix,adar2007user}. 
However, this makes data less useful~\cite{lakkaraju2008evaluating,kenneally2010dialing} and  prone to de-anonymization~\cite{coull2007playing}.
Other proposals require entities to send encrypted data to a central repository that aggregates contributions~\cite{pets10}. Locasto et al.~\cite{locasto2005towards} propose privacy-preserving data aggregation using Bloom filters, which, while constituting a one-way data structure, are vulnerable to simple guessing attacks. Secure distributed data aggregation is also discussed in~\cite{burkhart2010sepia,bilogrevic2014s}. While aggregation can help compute statistics, it only identifies most prolific attack sources and yields global models. As shown in~\cite{zhang2008highly}, however, generic attack models miss a significant number of attacks, especially when  sources choose targets strategically and focus on a few known vulnerable networks, thus yielding
poor prediction performance.

Previous work has also looked at the possible value of building collaborative and distributed intrusion detection systems. Katti et al.~\cite{katti2005collaborating} are among the first to study correlation among victims and demonstrated the prevalence of ``correlated'' attacks, i.e., attacks mounted by same sources against different victims. They find that: (1) Correlations among victims are persistent over time, and (2) Collaboration among victims from correlated attacks improves malicious IP detection time.
They also propose a collaboration mechanism in which victims learn from a centralized entity about other correlated victims, and can then query each other about ongoing attacks. 
Our work differs from~\cite{katti2005collaborating} as we introduce a controlled data sharing approach and study distributed collaborator selection strategies based on similarity measures, model different data sharing strategies, measure true and false positives of blacklisting recommendations, and address privacy concerns using efficient secure computation techniques.

\section{Preliminaries}\label{sec:preliminaries}

This section introduces notations and relevant background information.

\subsection{System Model}\label{sec:system}
In the rest of the paper, we assume a group of entities $\mathcal{V} = \{V_i\}_{i=1}^n$,
where each $V_i$ holds a dataset $L_i$ logging suspicious events,  such as, suspicious IP addresses observed by a firewall along with corresponding (time, port). 
For each $i$, we denote with $S_i$ the set of {\em unique} IP addresses in $L_i$.

Each entity $V_i$ aims to predict and blacklist IP addresses that will generate future attacks.
We consider a controlled data sharing model for collaborative predictive blacklisting, whereby entities estimate the benefits of collaborating in a privacy-preserving way, and then decide with whom, and what, to share. 
Each entity performs predictions based not only on its own dataset but also on an augmented dataset that comprises information shared by others, aiming to improve prediction and, at the same time, avoiding the wholesale disclosure of datasets.
To this end, we turn to efficient cryptographic protocols for privacy-preserving information sharing, presented below. %

\subsection{Cryptographic Tools}
\descr{Secure Two-Party Computation (2PC)~\cite{Yao}} allows two parties, on respective input $x$ and
$y$, to {\em privately} compute the output of any (public) function $f$ over
$(x,y)$. In other words, neither party learns anything about the counterpart's input beyond what can be inferred from $f(x,y)$. Security of 2PC protocols is formalized by considering an ideal implementation where a Trusted Third Party (TTP) receives the inputs and outputs the result of the function: Then, in the real implementation of the protocol (without a TTP), each party does not learn, provably, more information than in the ideal implementation. The first 2PC instantiation, based on garbled circuits, was presented by Yao~\cite{Yao} -- since then, optimizations and more efficient constructions have been presented, such as~\cite{HEKM11}.

\descr{Private Set Intersection (PSI)~\cite{FNP}} is a 2PC primitive 
that lets two parties, a server on input a set $S$, %
and a client on input a set $C$, %
interact so that the latter only learns $S\cap C$, and the former 
learns nothing (besides $|C|$).
State-of-the-art instantiations include both
garbled-circuit based techniques~\cite{HEK12,PSZ14} and specialized protocols~\cite{FNP,DT10,DT12}.

\descr{PSI with Data Transfer (PSI-DT)~\cite{DT10}} extends PSI as follows:
it involves a server on input a set $S$ where each item is associated to a data record,
and a client on input a set $C$. PSI-DT allows $C$ to learn 
the set intersection, along with the data records associated to the items in the
intersection (and nothing else), while $S$ learns nothing.

\descr{Private Set Intersection Cardinality (PSI-CA)~\cite{FNP,DGT12}} 
is a more ``stringent'' variant than PSI, as it only reveals the magnitude of the intersection,
but not the actual contents.

\descr{Private Jaccard Similarity (PJS)~\cite{espresso}} allows two parties, a server on input a set $S$, and a client on input a set $C$, to interact in such a way that 
the client only learns the Jaccard similarity~\cite{jaccard} between their respective sets, i.e.,~$J(\Sset,\Cset)=\frac{|S\cap C|}{|S\cup C|}=\frac{|S\cap C|}{|S|+|C|-|S\cap C|}$.
PJS can be instantiated using PSI-CA only, since secure computation techniques
(including PSI-CA) always reveal the size of inputs (i.e., size of sets in PSI-CA).

\smallskip In the rest of the paper, security of protocols discussed above is assumed in the honest-but-curious model, i.e., parties are assumed to follow protocol specifications and not to misrepresent their inputs, but, during or after protocol execution, they might attempt to infer additional information about other parties' inputs.

\subsection{Predictive Blacklisting}\label{subsec:predictive}

Let $t$ denote the day an attack was reported and $T$ the current time, so $t = 1, 2, ..., T$. We define a training window, $T_\text{\em train}$ and a testing window, $T_\text{\em test}$. Prediction algorithms usually rely on information in the training data, $t \in T_\text{\em train}$, to tune their model and validate the predictions for the testing data, $t \in T_\text{\em test}$.

The Global Worst Offender List (GWOL) is a basic prediction algorithm that selects top attack sources from $T_\text{\em train}$, i.e., highest number of globally reported attacks~\cite{zhang2008highly}. 
Local Worst Offender List (LWOL) is the local version of GWOL and operates on a local network based entirely on its own history~\cite{zhang2008highly}. 
LWOL fails to predict on attackers not previously seen, while GWOL tends to be irrelevant to small victims. 
Thus, machine learning algorithms were suggested to improve GWOL and LWOL~\cite{soldo,zhang2008highly}. 

We use the {\em Exponentially Weighted Moving Average (EWMA)} algorithm, as proposed by Soldo et al.~\cite{soldo}, to perform blacklisting prediction. EWMA uses time series aggregation: It aggregates attack events from $T_\text{\em train}$ to predict future attacks. 
Note that it is out of the scope of this paper to improve on existing prediction algorithms. Rather, we focus on evaluating the feasibility of controlled data sharing for collaborative threat mitigation and, specifically, on measuring how different collaboration strategies perform in comparison to each other.

\descr{Accuracy Metrics.} As commonly done with prediction algorithms, we measure accuracy with {\bf {\em True Positives} (TP)}, which is the number of predictions that correctly match future events. 
In practice, potentially malicious sources might not be blacklisted at once as blacklisting algorithms rely on several observations over time, such as the rate at which the source is attacking or the payload of suspicious packets. Therefore, it is important to distinguish between the {\em prediction} and the {\em blacklisting} algorithm:
the former identifies potential malicious sources and/or creates a watch-list, which is fed to the latter in order to help decide whether or not to block sources.
The prediction algorithm thus enables the identification of suspicious IP addresses that deserve further scrutiny and improve the effectiveness of blacklisting algorithms. 
Therefore, prior work~\cite{soldo,zhang2008highly} focused almost exclusively on measuring TP and ignored other accuracy measures such as false positives. By contrast, we decide to also study \textbf{\emph{False Positives (FP)}}, i.e., the number of predictions that are incorrect. This measurement helps us better understand the possible negative overhead introduced by data sharing.

\descr{Upper Bounds.} As in previous work~\cite{soldo}, we use two upper bounds to evaluate the accuracy of the prediction, aiming to take into account the fact that a future attack can be predicted only if it already appeared in the logs of some victims. The Global Upper Bound $\mbox{GUB}(V_i)$ measures, for every target $V_i$, the number of attackers that are both in the training window \emph{of any victim} and in $V_i$'s testing window. For every $V_i$, we also define the Local Upper Bound $\mbox{LUB}(V_i)$, as the number of attackers that are both in $V_i$'s training and testing windows.

\section{Collaborative Predictive Blacklisting via Controlled Data Sharing}\label{sec:framework}

We outline our controlled data sharing approach for collaborative predictive blacklisting.
It involves three steps: %
\begin{enumerate}
\item Estimating the benefits of sharing security data between potential partners, in a privacy-preserving way (i.e., without disclosing the datasets);
\item Establishing partnerships;
\item Sharing data in a privacy-respecting way and guaranteeing that collaborating entities only share what is agreed upon.
\end{enumerate}

\begin{table}[b]
\centering
\begin{minipage}[b]{0.48\hsize}
\resizebox{1.025\linewidth}{!}{
\begin{tabular}{ | c | c | c | }
  \hline                        
  {\bf Benefit Esti-} & {\bf Operation}  & {\bf Private}  \\
  {\bf mation Metric} & & {\bf Protocol}\\
  \hline                        
  \emph{Intersection-} & \multirow{2}[2]{*}{$|S_i\cap S_j|$}  & \multirow{2}[2]{*}{PSI-CA~\cite{DGT12}}\rule{0pt}{2ex}\\
  \emph{Size} & &\\[0.5ex]
  \hline
  \emph{Jaccard} & $\dfrac{|S_i \cap S_j|}{|S_i \cup S_j|}$ & PJS~\cite{espresso}
  \rule{0pt}{4ex}\rule[-0.9ex]{0pt}{0pt}\\[2ex]
  \hline
  \multirow{2}[2]{*}{\emph{Pearson}} & \multirow{2}[2]{*}{$\sum_{l=1}^{N}\dfrac{(s_{i_l} - \mu_i)(s_{j_l}-\mu_j)}{N \sigma_i \sigma_j}$} & Garbled\rule{0pt}{2.5ex}\\
  & & Circuits~\cite{HEKM11}
  \rule[-0.9ex]{0pt}{0pt}\\[1ex]
  \hline    
  \multirow{2}[2]{*}{\emph{Cosine}} & \multirow{2}[2]{*}{$\dfrac{\vec{S_i}\vec{S_j}}{\|\vec{S_i}\| \|\vec{S_j}\|}$} & Garbled\rule{0pt}{2.5ex}\\
  & & Circuits~\cite{HEKM11}
  \rule[-0.9ex]{0pt}{0pt}\\[1ex]
  \hline    
\end{tabular}
}
\vspace{-0.1cm}
\caption{Metrics for estimating  benefits of collaboration between $V_i$ and $V_j$, along with corresponding protocols for their secure computation. $\mu_i,\mu_j$ and $\sigma_i,\sigma_j$ denote, resp., mean and standard deviation of $\vec{S_i}$ and $\vec{S_j}$.}
\label{tab:metric-1}
\end{minipage}
\hfill
\begin{minipage}[t]{0.48\hsize}
\vspace{-5.16cm}
\resizebox{1.025\linewidth}{!}{
\begin{tabular}{ | c | c | c | }
  \hline                        
  {\bf Sharing Strategy} & {\bf Operation}  & {\bf Private Protocol}  \\
  \hline                        
  \emph{Intersection} & $S_i\cap S_j$ & ~PSI~\cite{DT10}
    \rule{0pt}{2.5ex}\rule[-0.9ex]{0pt}{0pt}\\[1ex]
  \hline  
  \emph{Intersection with} & $\{(\mbox{IP,time,port})|$ & \multirow{2}[2]{*}{PSI-DT~\cite{DT10}}\rule{0pt}{3ex}\\
  \emph{Associated Data} & $\mbox{IP}\in S_i\cap S_j\} $ & \\[1ex]
  \hline  
  \emph{Union with} & $\{(\mbox{IP,time,port})|$ & \multirow{2}[2]{*}{No Privacy}\rule{0pt}{3ex}\\
  \emph{Associated Data} & $\mbox{IP}\in S_i\cup S_j\} $ & \\[1ex]
  \hline
\end{tabular}
}
\vspace{-0.1cm}
\caption{Strategies for data sharing among partners $V_i$ and $V_j$, along with corresponding protocols for their secure computation.}
\label{tab:metric-3}
\end{minipage}
\vspace{-0.2cm}
\end{table}

\subsection{Benefit Estimation}\label{subsec:select}

We consider several similarity metrics to estimate the benefits of collaboration: We report them
in Table~\ref{tab:metric-1}, along with the corresponding protocols for their privacy-preserving computation.
We look at similarity metrics since previous work~\cite{katti2005collaborating, zhang2008highly} has shown that collaborating with {\em correlated victims} works well, i.e., entities targeted by attacks mounted by the same source against different networks (around the same time). Intuitively, correlation arises from attack trends as correlated victim sites might be on a single hit list or natural targets of a particular exploit. 

We consider two set-based metrics, i.e., \emph{Intersection-Size} and \emph{Jaccard}, which measure set similarity and operate on unordered sets, as well as \emph{Pearson} and \emph{Cosine} similarity, which provide a more refined measure of similarity as they also capture statistical relationships. 
The last two metrics operate on data structures representing attack events, such as a binary vector, e.g., $\vec{S_i}=[s_{i_1}~s_{i_2}\cdots s_{i_N}]$, of all possible IP 
addresses with 1-s if an IP attacked at least once and 0-s otherwise. This can make
it difficult to privately compute correlation in practice, as both parties need to agree on the
range of IP addresses under consideration to construct vector $\vec{S_i}$.
Considering the entire range of IP addresses is not reasonable (i.e., this would
require a vector of size 3.7 billion, one entry for each routable IP address).
Instead, parties could either agree on a range via secure computation or fetch predefined ranges from a public repository. 

All the functions we consider are symmetric, i.e., both parties obtain the same value, however,
some of the protocols used for secure computation of the benefit estimate, such as PSI-CA~\cite{DGT12} and PJS~\cite{espresso}, reveal the output of the protocol to only one party. Without loss of generality, we assume that this party always reports the output to its counterpart, which is a common assumption in the honest-but-curious model.

In practice, one could use any combination of metrics. Also, the list in Table~\ref{tab:metric-1} is non-exhaustive and other metrics could be considered, 
as long as it is possible to efficiently support their privacy-preserving computation.

\subsection{Establishing Partnerships}\label{subset:partner}
After estimating the benefits of collaboration, in order to establish partnerships,
entities could follow different strategies, acting in a distributed way or relying on a coordinating entity.
For instance, an organization could request the collaboration of all entities for which estimated benefits are above a threshold (i.e., based on a {\em ``local threshold''}), or enlist the $k$ partners with maximum expected benefits ({\em ``local maximization''}).
Local approaches have the advantage of not involving any third parties, but may require complex negotiations in order to reach a partnership agreement, as collaboration incentives may be asymmetric, e.g., party A might be willing to collaborate with B, but B might prefer to do so with C. 
With centralized approaches, a semi-trusted server collects estimated benefits (but not datasets) and clusters entities so that those in the same cluster collaborate ({\em ``clustering''}), or encourage sharing among the pairs with highest expected benefits seeking a global utility-vs-cost optimum ({\em ``global maximization''}).

Naturally, an appropriate partner selection strategy heavily depends on the use-case scenario and the trade-offs that organizations are willing to pursue.  
Hence, some strategies might work well in different settings depending on economic, strategic, and operational factors. The evaluation of the different partnership strategies is an interesting research problem, particularly amenable to a game-theoretic analysis.
In this work, we {\em do not experiment with different strategies to establish partnerships} and leave such an analysis for future work. As a result, in our experiments (Section~\ref{sec:experiments}), we fix one partner selection strategy and focus on the evaluation of different benefit estimation and data sharing mechanisms.

\subsection{Data Sharing}\label{subsec:merge}
After two entities have established a partnership, they can proceed to share their data with each other.
This process can occur in different ways, e.g., they can disclose their whole datasets or only share which IP addresses they have in common, with or without all attack events associated to common addresses. %

Following our controlled data sharing approach, nothing is to be disclosed beyond what is agreed upon (and, ideally, what is beneficial). For instance, if partners agree to only share information about common attackers, they should not learn any other information. 
Possible sharing strategies we consider, along with the corresponding privacy-preserving protocols, are reported in Table~\ref{tab:metric-3}. Again, we assume that the output of the sharing protocol is revealed to both parties.

Strategies denoted as \emph{Intersection/Union with Associated Data} mean that 
parties not only compute and share the intersection (resp., union), but also all events related to items in the resulting set.
Obviously, Union with Associated Data does not yield any privacy, as all events are mutually shared, but we include it to compare its efficacy to Intersection with Associated Data.

\section{The DShield Dataset}\label{sec:dshield}
As we aim to evaluate the viability of the controlled data sharing
approach and compare how different sharing strategies impact prediction accuracy,
we need to design an experiment involving real-world data pertaining to suspicious IP addresses
and observed by different organizations. To this end, as done in prior work~\cite{katti2005collaborating,soldo,zhang2008highly}, we turn to Dshield.org~\cite{dshield}:
this section introduces the data we collect from Dshield, and the methodology we use
to clean it and to design a meaningful data sharing experiment.

\begin{figure*}[t]
\centering
 \subfigure[]{
   \includegraphics[trim=0 0 0 0, scale =0.32] {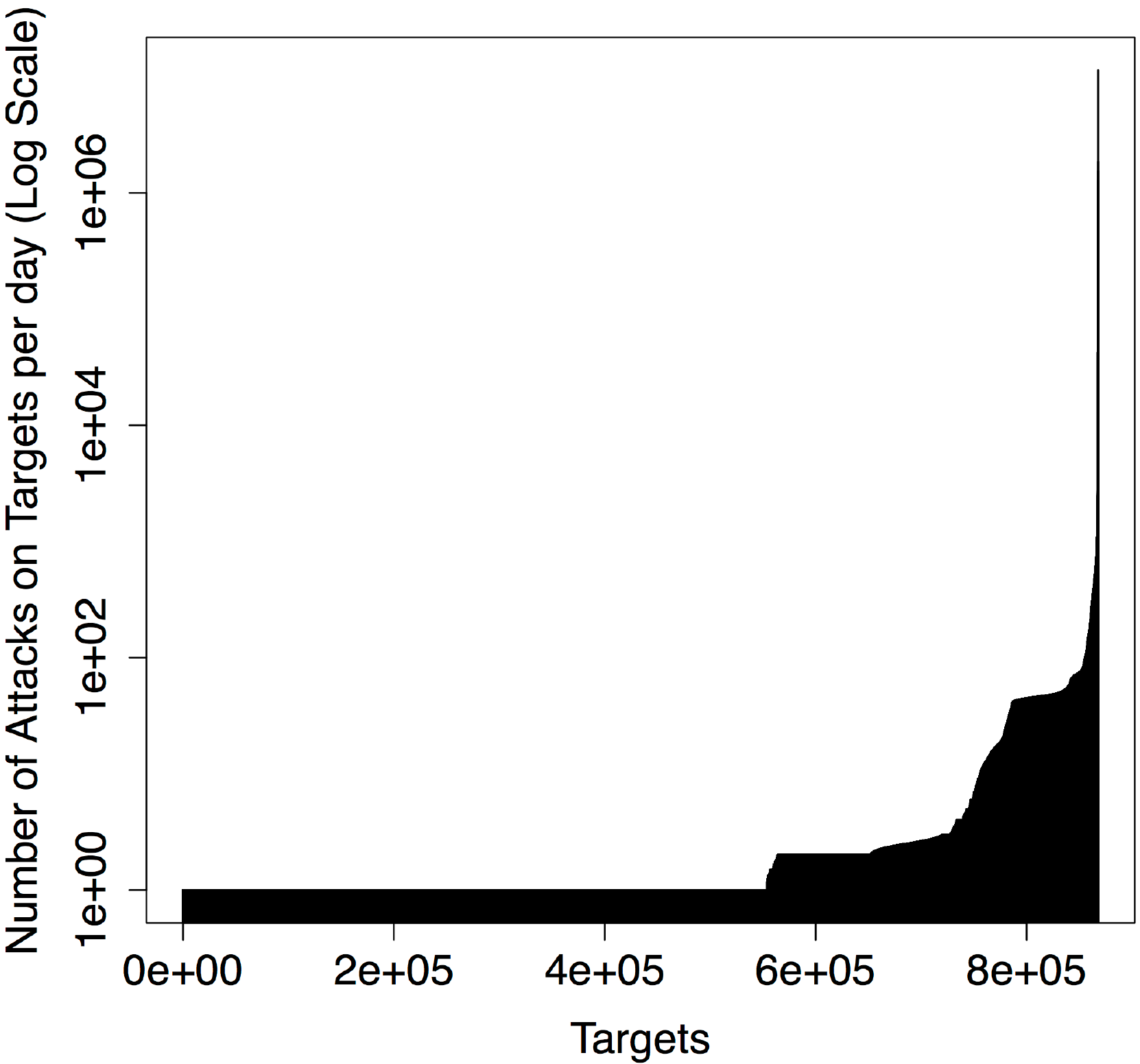}
   \label{fig:average:subfig1}
 }
 \hspace{-0.2cm}
\subfigure[]{
   \includegraphics[trim=0 0 0 0, scale =0.32] {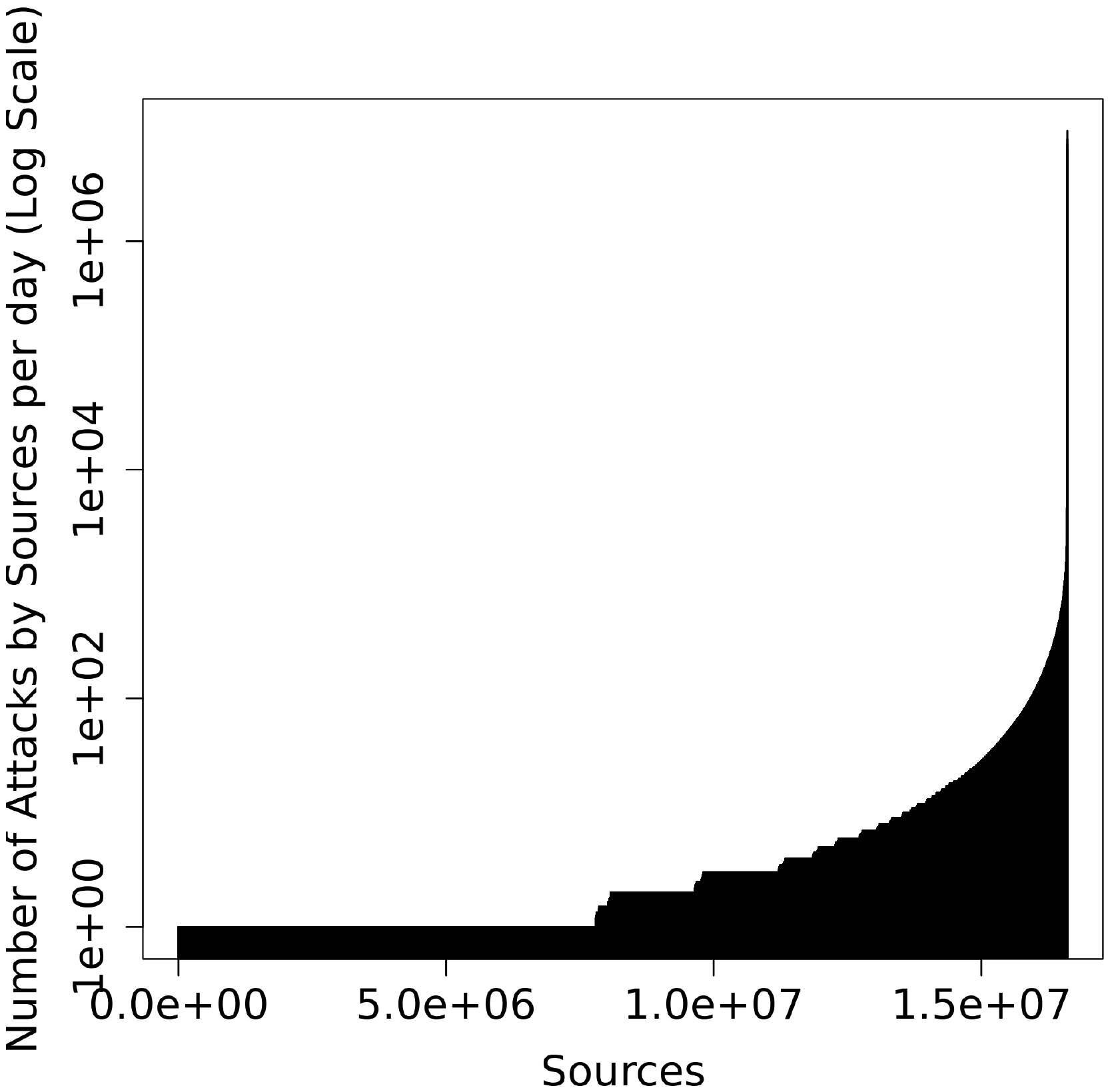}
   \label{fig:average:subfig2}
 }
 \\[-1.6ex]
\subfigure[]{
   \includegraphics[trim=0 0 0 0, scale =0.32] {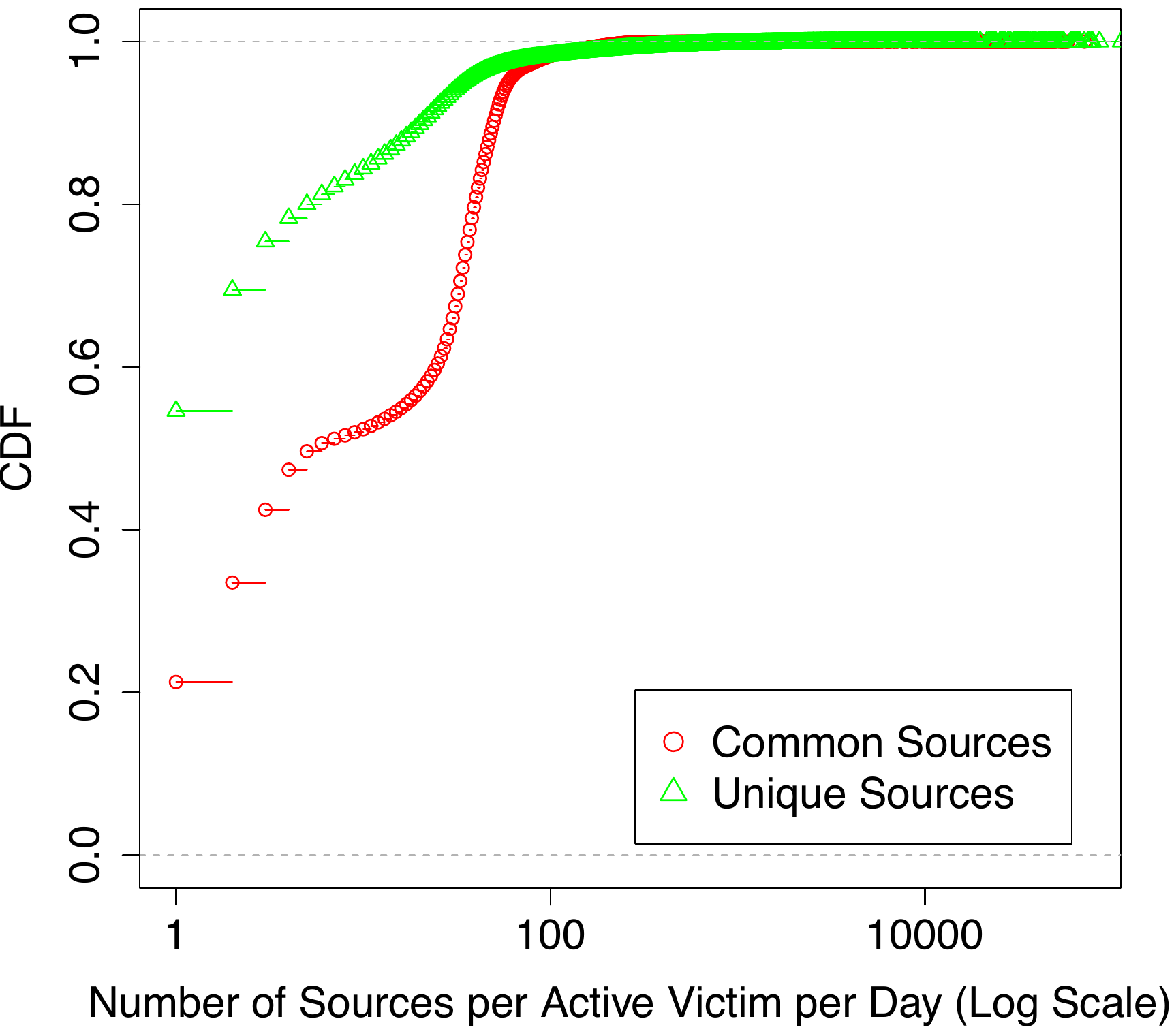}
   \label{fig:average:subfig3}
 }
 \hspace{-0.27cm}
 \subfigure[]{
   \includegraphics[trim=0 0 0 0, scale =0.32] {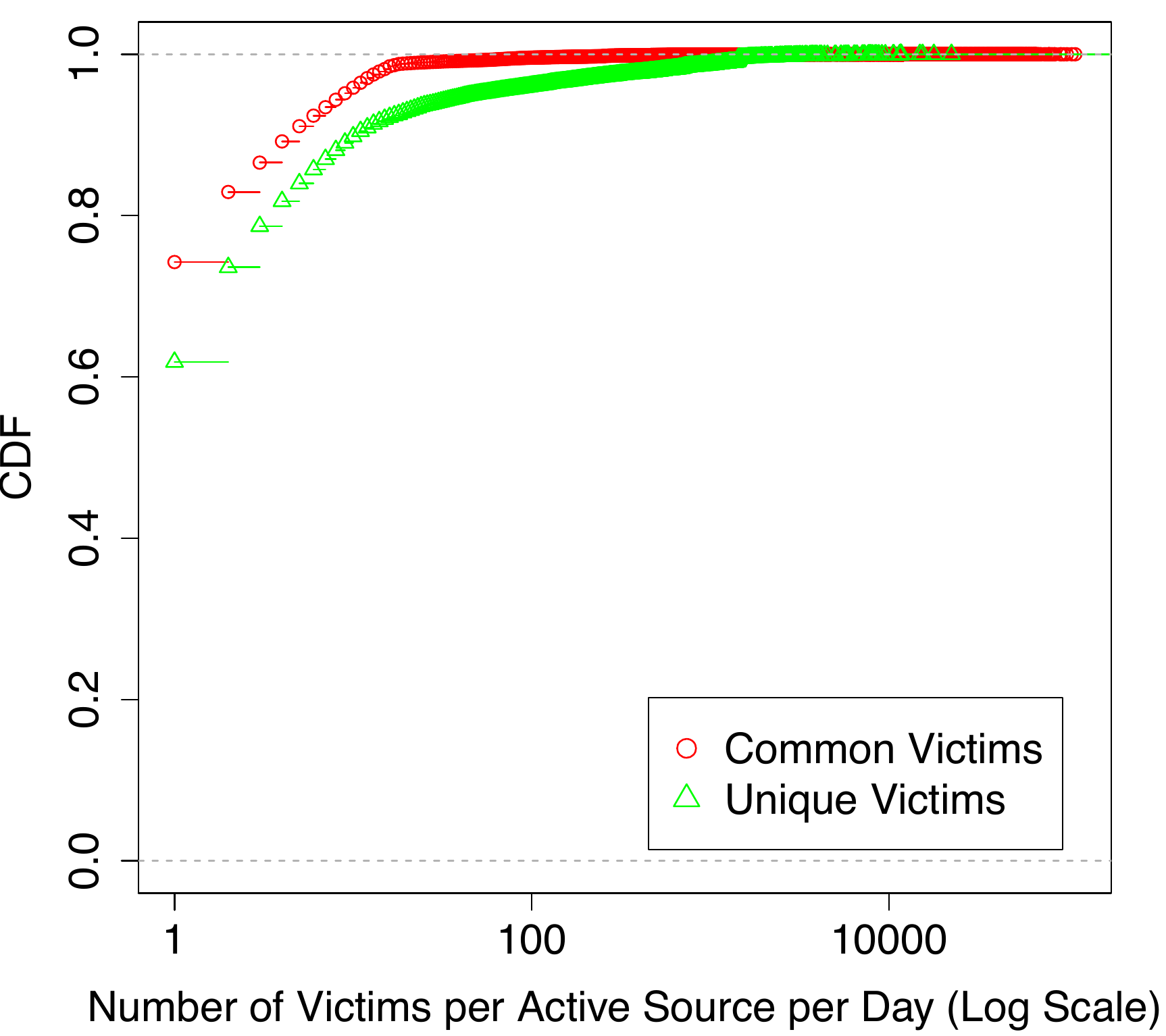}
   \label{fig:average:subfig4}
   \hspace{-0.3cm}
 }

\label{fig:average}
\vspace{-0.3cm}
\caption{Number of attacks per day: (a) On all targets, and (b) by all sources. CDF of the daily number of common and unique: (c) Sources per active victims, and (d) victims per active sources.} 

\end{figure*}

\subsection{Original Dataset}  
We obtained two months' worth of anonymized logs from Dshield.org~\cite{dshield},
a community based repository of intrusion detection system
logs %
that publishes blacklists of most prolific attack sources reported in these logs.
Each entry in the logs includes an anonymized Contributor ID (the target), a source IP address (the suspected attacker), a target port number, and a timestamp, as illustrated in Table~\ref{tab:illustrationDShield}. 
Note that DShield anonymized the ``Contributor ID'' field by replacing it with 
a random yet {\em unique} string that maps to a single victim.

\begin{table}[ttt]
\small
\centering
\begin{tabular}{ | c | c | c | c | c | }
  \hline                        
  {\bf Contributor ID} & {\bf Source IP}  & {\bf Target port} & {\bf Timestamp} \\
  \hline                        
  44cc551a & 211.144.119.042 & 1433 &  2013-01-01 11:48:36 \\
  \hline  
\end{tabular}
\vspace{0.1cm}
\caption{Example of an entry in the DShield logs.}
\label{tab:illustrationDShield}
\vspace{-0.5cm}
\end{table}

The data collected from DShield consists of about 2 billion entries, generated by $800$K unique contributors, including more than $16$M malicious IP sources, for a total of $170$GB worth of logs. 
We pre-processed the dataset in order to reduce noise and erroneous entries, following the same methodology adopted by previous work on DShield data~\cite{soldo,zhang2008highly}: We removed approximately 1\% of of all entries, which belonged to invalid, non-routable, or unassigned IP addresses, or referred to non-existent port numbers. 

\subsection{Measurements \& Observations} 
We performed a small measurement analysis of our DShield dataset, aiming to better understand characteristics of attackers and victims. 
Overall, our observations are in line with prior work~\cite{pouget2005vh,soldo} and demonstrate that attackers tend to hit victims in a coordinated fashion, thus confirming the potential for collaboration.

\descr{General Statistics.}
We observe that $75\%$ of targets contribute less than $10\%$ of the time, while $6\%$ of targets ($50,000$ targets) contribute daily. We describe, at the end of this section how we filter out targets that seldom contribute. For more details and statistics, we refer to the Appendix.

\descr{Victims' Profile.}
Fig.~\ref{fig:average:subfig1} shows the number of attacks per day on targets, with mean number of daily attacks on targets of $58.46$ and median of $1$. We observe three distinct victims' profiles: (1) Rarely attacked victims: $87\%$ of targets get less than $10$ attacks day, indicating many victims seldom attacked; (2) Lightly attacked victims: $11\%$ of victims get $10$ to $100$ attacks a day; (3) Heavily attacked victims: Only $2\%$ of targets are under high attack (peaking at $11$M a day). In other words, most attacks target few victims. 

\descr{Attackers' Profile.} Fig.~\ref{fig:average:subfig2} shows the number of victims attacked by each source per day, with mean number of daily attacks of $45.85$ and median of $2$. We observe that $80\%$ of sources initiate less than $10$ attacks a day. A small number of sources generates most attacks (up to $10$M daily). This  indicates two main categories of attackers: Stealth and heavy hitters. In our data set, we observe that several of top heavy attackers (more than $20$M attacks) come from IP addresses owned by ISPs in the UK.

\descr{Attacks' Characteristics.} Fig.~\ref{fig:average:subfig3} shows the Cumulative Distribution Function (CDF) of the number of unique sources seen by each {\em active} target a day. We focus on active victims: Victims that did report an event on that particular day because, as previously discussed, many victims report attacks rarely thus creating a strong bias towards $0$ otherwise. The figure contains attackers shared with other targets (common attackers) and attackers unique to a specific victim. 90\% of victims are attacked by at most $40$ unique sources and $60$ shared sources. This shows that, from the victim's perspective, targets observe more shared sources than unique ones. Compared to previous work~\cite{soldo, katti2005collaborating}, this reinforces the past trend of targets having many common attackers. 
Fig.~\ref{fig:average:subfig4} shows that $90\%$ of sources attack $30$ common victims and $60$ unique victims. Although attackers share a large number of common victims, they also attack other victims. In Fig.~\ref{fig:average:subfig3} and Fig.~\ref{fig:average:subfig4}, we observe again  three types of victims and two types of attackers.

\descr{Observations.} A significant proportion of victims ($\sim$70\%) contributes a single event overall. After thorough investigation, we find that these \emph{one-time contributors} can be grouped into clusters all reporting the same IP address within close time intervals (often within one second). Many contributors share only one attack event, at the same time, about the same potentially malicious IP address. 
Similarly, many contributors only contribute one day out of the two months. 
These contributors correlate with the aforementioned one-time contributors. 

\subsection{Final Dataset}
In order to select a meaningful dataset for our experiments,
we remove contributors that do not report much information. Specifically, we remove victims that contribute either (1) only one event overall, or (2) only one day and less than 20 events over the two-month period. 
This reduces the number of considered victims from $800$,$000$ to $188$,$522$, resulting in the removal of about 2 million attacks.
This filtering maintains a high diversity of contributors, and seeks to model real-world scenarios (as opposed to focusing on large contributors only).

In summary, our final dataset includes 2 billion attacks, contributed by almost $190K$ entities  over 60 days, each reporting an average of $200$ suspicious (unique) IPs and $2$,$000$ attack events.

\section{Experimental Analysis}\label{sec:experiments}

We compare different benefit estimation metrics and sharing strategies, by measuring improvements to prediction accuracy, using the final Dshield dataset.
The dataset and the source code (written in R) used in our experiments are available upon request.

\subsection{Experimental Setup}
Our objective is to design an experiment that is easy to reproduce and enables a meaningful evaluation of controlled data sharing. We describe below our experimental setup and introduce our modeling assumptions. 

\descr{Sampling of Potential Partners.} Our final dataset includes $188$,$522$ victims. In theory, we could evaluate the performance of controlled data sharing by considering all possible collaboration pairs. However, this would be impractical. As a result, we follow a {\em sampling} approach, i.e., we select $100$ entities at random from all possible victims. We then evaluate different collaboration strategies considering the $100\cdot99/2 =4$,$950$ possible pairs, and average results over $100$ independent iterations.
The random sampling model is {\em ``conservative''} in the resulting improvement of the prediction accuracy, as it is likely to do worse than if entities considered non-random potential partners, e.g., organizations in the same sector.

\descr{Benefit Estimation and Partner Selection.} We consider four privacy-preserving metrics for estimating the benefits of sharing (as discussed in Section~\ref{subsec:select}): {\em Jaccard}, {\em Pearson}, and {\em Cosine} similarity and {\em Intersection-Size}.
Each metric is computed pairwise: 
For each metric, we obtain a $100 \times 100$ matrix estimating data sharing benefits among all possible pairs of organizations. 

Recall from Section \ref{subset:partner} that, while we mention a few possible strategies to select partners,
we do not evaluate them in this paper, as such mechanisms are out of scope. 
For simplicity, we consider an approach similar to a \emph{global maximization}: We partner entities with the highest values in the similarity matrix. Specifically, we select the top 1\% collaboration pairs (i.e., $50$ pairs) with the maximum expected benefits. This is likely a conservative stance as we consider only a small number of partnerships (i.e., only few entities collaborate).
This approach results in some entities sharing data with several partners, and others not collaborating with anyone. We define the {\em number of collaborators} as the number of distinct entities (out of $100$) that are selected in the $50$ collaboration pairs. 
We also define the {\em coalition size} as the number of other entities an organization collaborates with.

\begin{figure*}[t!]
\centering
 \subfigure[]{
   \includegraphics[trim=0 0 0 0, scale=0.32]{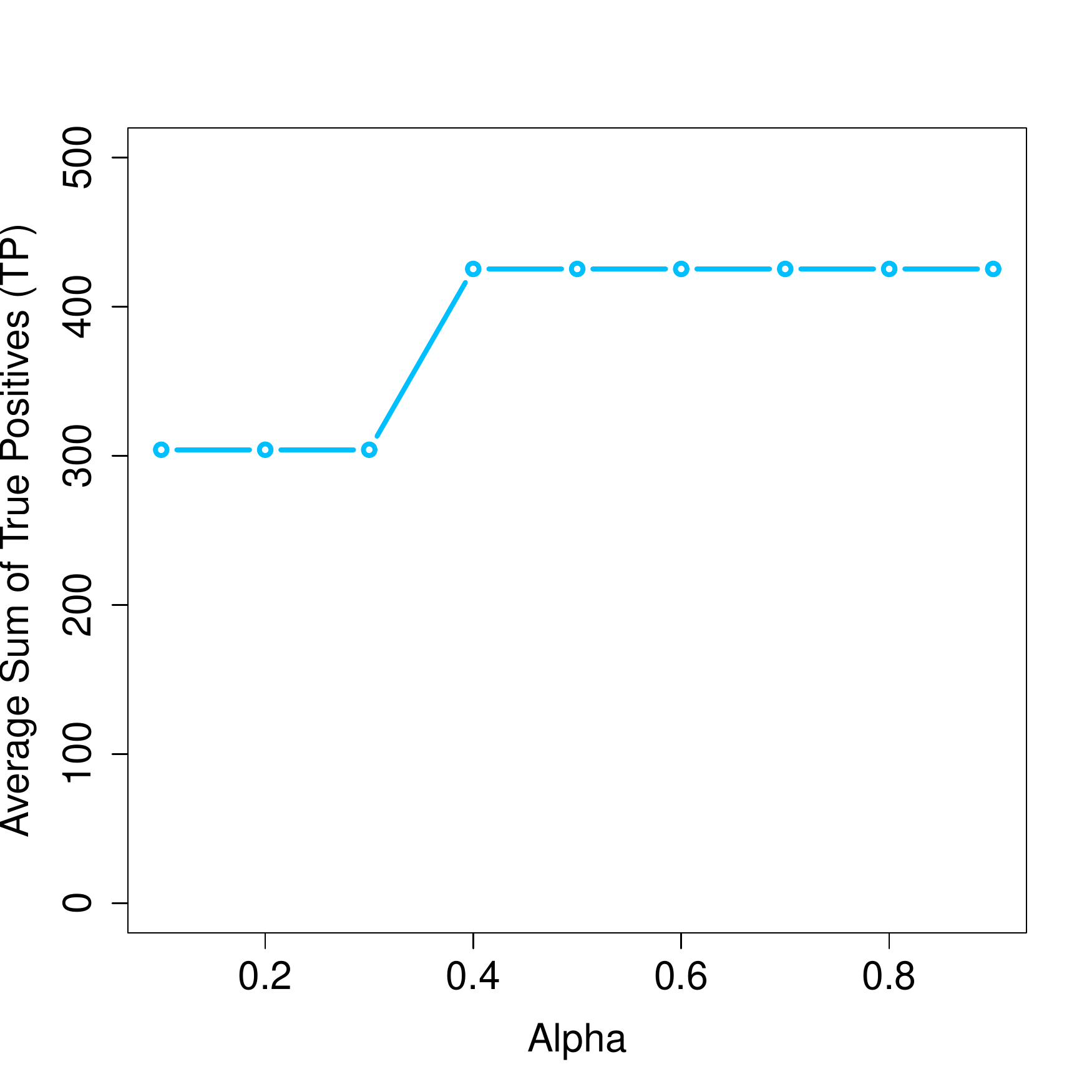}
   \label{fig:prediction:alpha}
}
 \subfigure[]{
   \includegraphics[trim=0 0 0 0, scale =0.32] {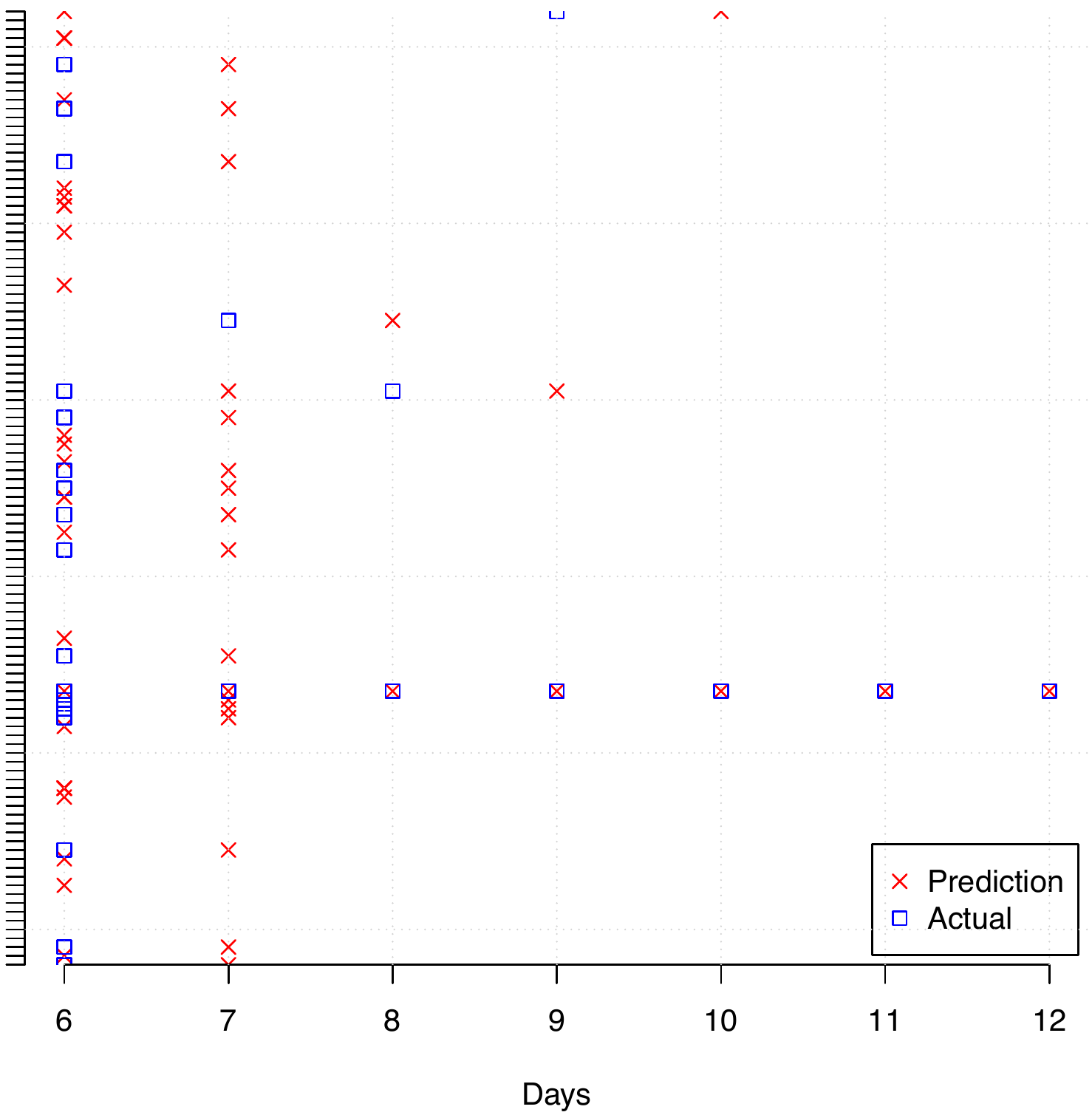}
   \label{fig:prediction:visual}
 }
   \vspace{-0.3cm}
\label{fig:alpha}
\caption{Evaluation of baseline prediction (with no collaboration). (a) Number of true positives for different values of prediction algorithm parameter $\alpha$. (b) Visualization of a victim's predictions over time for a series of attackers with $\alpha=0.9$ on y-axis.
}
\end{figure*}

\descr{Sharing.} As described in Section~\ref{subsec:merge}, we consider three types of data sharing strategies,  (1) {\em Intersection}, 
(2) \emph{Union with Associated Data}, and (3) \emph{Intersection with Associated Data}.
Since (1) is likely to yield poor results, we do not consider it in our experiments.
With (2), partners share all data known by each party prior to current time $t$: It is a generous strategy that enriches others' datasets rapidly. Whereas, with (3), partners only share events from those IP addresses that attacked both partners (i.e., the intersection). 

\descr{Prediction.} We use a five day window to train our prediction algorithm ($T_\text{\em train}=5$) and aim to predict attacks for the next day ($T_\text{\em test}=1$). Although our dataset contains two month worth of data, in order to speed up our experiments, we focus our analysis over a one-week period, i.e., we predict attacks on days 6 to 12, using the previous five days as the training dataset.

\descr{Accuracy.} 
As anticipated in Section~\ref{subsec:predictive}, we measure the prediction success by computing the number of True Positives (TP), similar to prior work~\cite{soldo,zhang2008highly}.
True positives correspond to successfully predicted attacks. We measure prediction improvement as:\vspace{-0.2cm} $$I = (\mbox{TP}_{c}-\mbox{TP})/\mbox{TP},\vspace{-0.2cm}$$
where $\mbox{TP}$ is the number of true positives before collaboration and $\mbox{TP}_c$ is the number of true positives after collaboration. 
In the following, we give both improvement measures over all entities, and for collaborating entities only. 
Unlike previous work~\cite{soldo,zhang2008highly}, we also measure False Positives (FP) aiming to measure the potential negative effects of controlled data sharing. This allows us to take precautions on the notion of data sharing and more effectively compare different benefit estimation and data sharing strategies.

\subsection{Different Benefit Estimation Metrics}

\begin{figure*}[t!]
\centering
 \subfigure[]{
   \includegraphics[trim=0 0 0 0, scale =0.3] {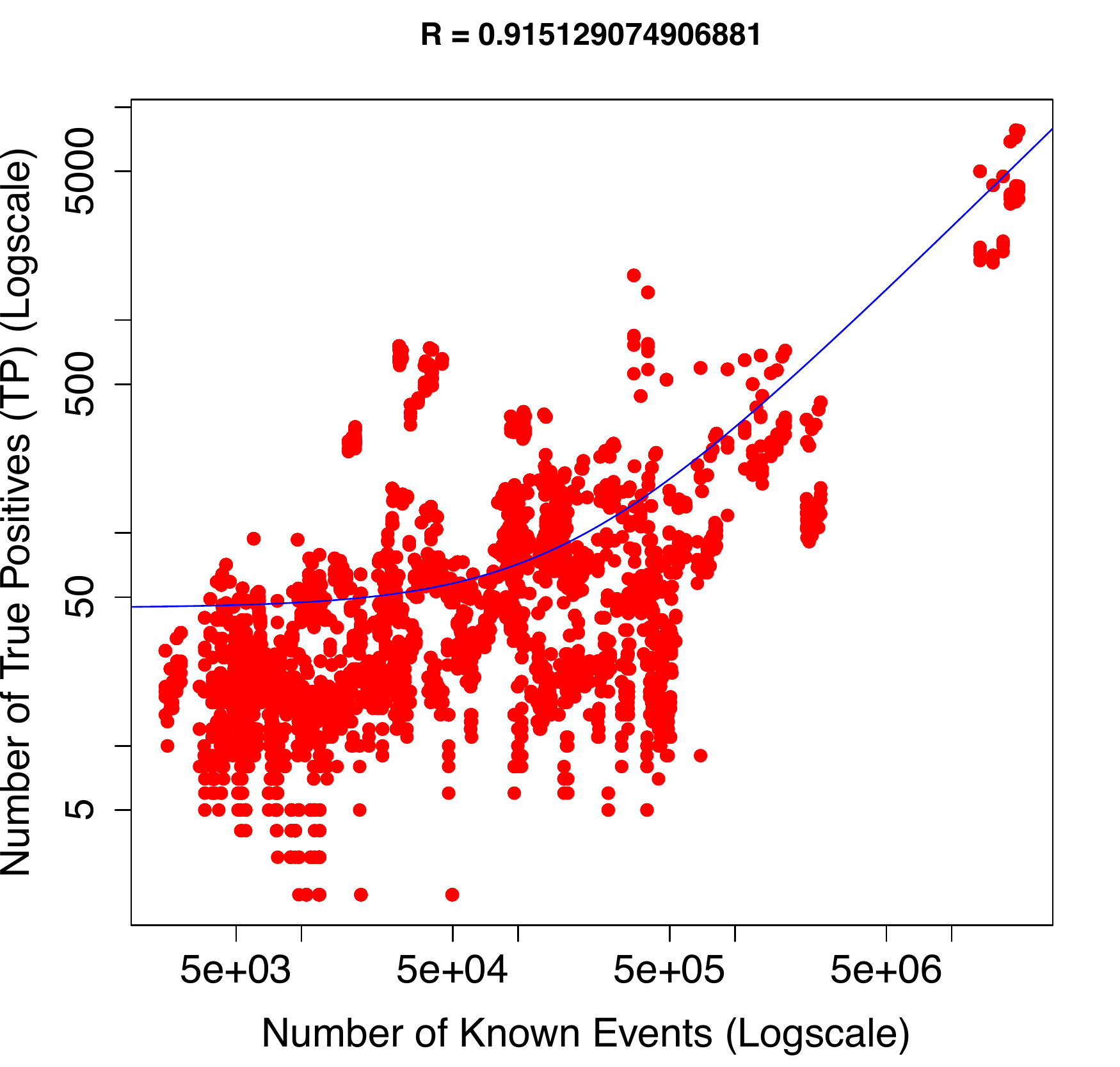}
   \label{fig:collaboration:subfig1}
 }
  \hspace{-0.25cm}
\subfigure[]{
   \includegraphics[trim=0 0 0 0, scale =0.3] {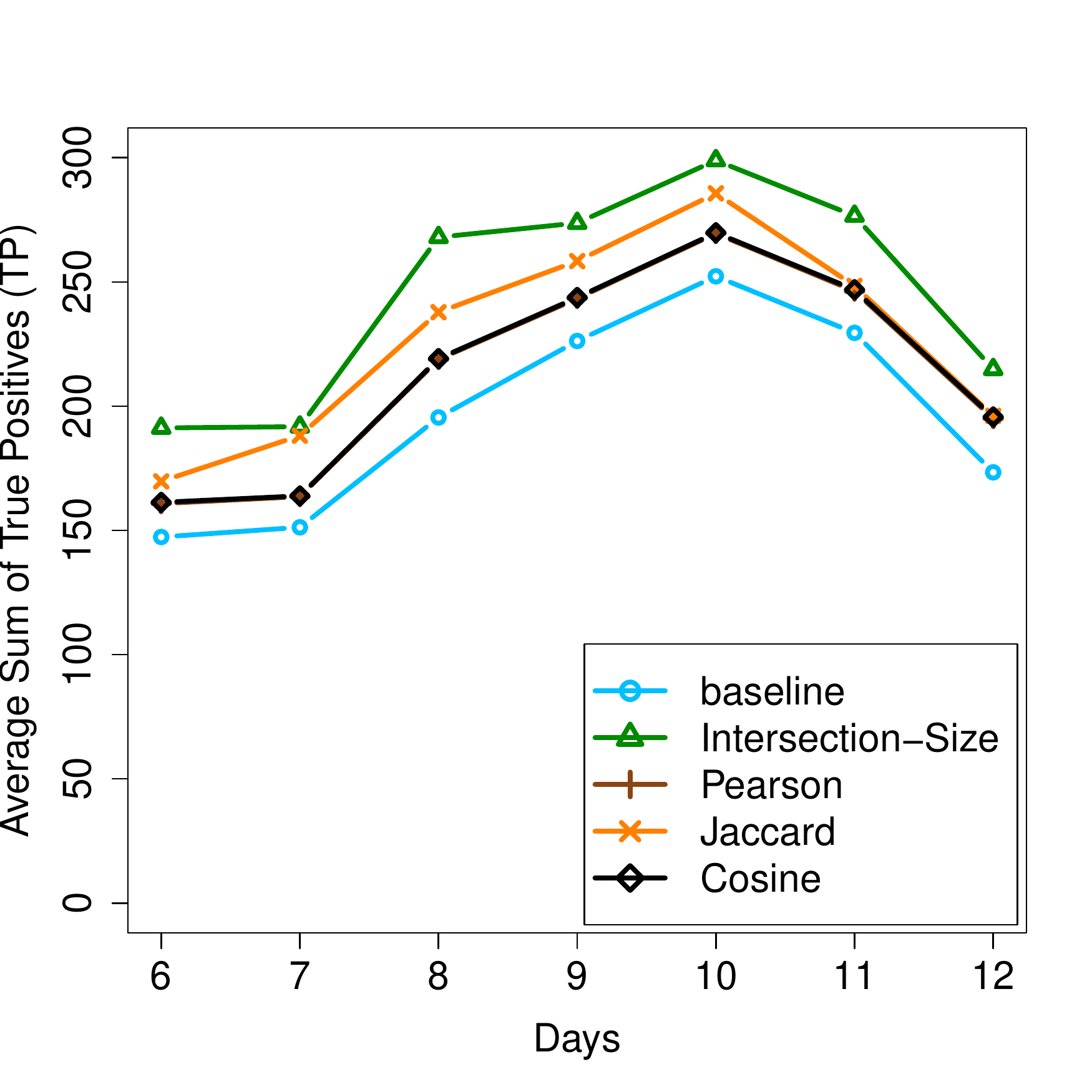}
   \label{fig:collaboration:subfig2}
 }
  \hspace{-0.25cm}
 \subfigure[]{
   \includegraphics[trim=0 0 0 0, scale =0.3] {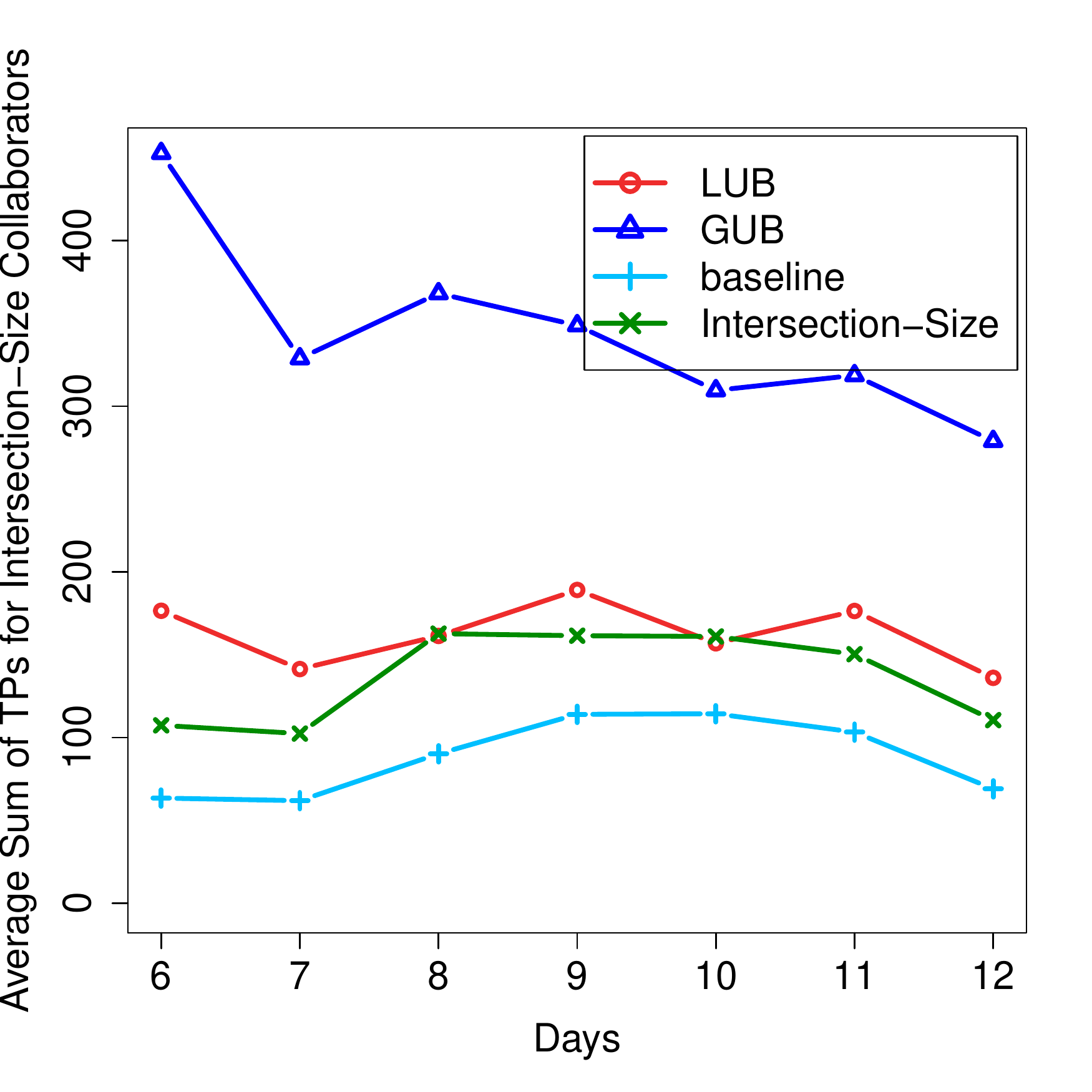}
   \label{fig:collaboration:subfig3}
    }
\label{fig:collaboration}
\vspace{-0.3cm}
\caption{Prediction Analysis. (a) Correlation between number of events known by targets, and their ability to predict attacks. The blue curve shows the linear regression (note the log-log scale). (b) Average sum of True Positives over time for different benefit estimation methods. (c) Average sum of True Positives over time among collaborators selected by \emph{Intersection-Size} including upper bounds (LUB and GUB). 
}
\vspace{-0.1cm}
\end{figure*}

\descr{Determining the Value of $\alpha$.} 
Before testing the performance of different strategies, we need to identify appropriate $\alpha$ values for the EWMA prediction algorithm by evaluating the performance of the prediction. 
For small values of $\alpha$, the prediction algorithm aggregates past information uniformly across the training window to craft predictions. In other words, events in the distant past have a similar weight to events in the recent past and the algorithm has a long memory. On the contrary, with a large $\alpha$, the prediction algorithm focuses on events in the recent past. %
Fig.~\ref{fig:prediction:alpha} shows the evolution of the baseline  prediction for different values of $\alpha$, plotting the True Positives (TP) sum of all $100$ victims averaged over $100$ iterations. Values between $\alpha=0.4$ and $\alpha=0.9$ perform best. This can be explained by remembering the ``bursty nature'' of web attacks, as discussed in Section~\ref{sec:dshield}.  
As a result, we set $\alpha = 0.9$.

\descr{Visualizing Predictions.} Fig.~\ref{fig:prediction:visual} shows a visualization of the prediction. When an attack occurs (blue square), the algorithm systematically predicts an attack (red cross) in the next time slot. Because $\alpha=0.9$, the last attack event has a larger weight.

\descr{Baseline Prediction.}
We verify the effectiveness of the prediction algorithm by correlating the information known prior to collaboration with the ability to predict attacks. As expected, targets that know more about past attacks (large $S_i$), successfully predict more future attacks. We measure correlation $R>0.9$ on average, which indicates strong correlation between knowledge and prediction. This suggests that collaboration helps prediction. We visualize the correlation between knowledge and prediction accuracy for all victims in our final dataset running the EWMA prediction in Fig.~\ref{fig:collaboration:subfig1}.

\descr{Benefit Estimation Strategies.}
Fig.~\ref{fig:collaboration:subfig2} illustrates the accuracy of predictions for different benefit estimation strategies over the course of one week, {\em fixing the sharing strategy to Intersection with Associated Data}, as it is more conservative than sharing everything (i.e., the union). We sum the total number of TP for both ``collaborators'' (i.e., entities that do share data) and ``non-collaborators'' (entities that do not share data, thus performing as in the baseline). We observe that \emph{Intersection-Size} performs best, followed by \emph{Jaccard},
and \emph{Cosine/Pearson}. 
The overall decrease in sum of true positives after day $10$ is due to less attacks 
reported in those days (see Fig.~\ref{fig:general:subfig1}). 

\descr{Improvement Over Baseline.} In Fig.~\ref{fig:collaboration:subfig3}, we compare the prediction accuracy of the upper bounds, the baseline, and collaboration using \emph{Intersection-Size} for benefit estimation (again, while sharing using \emph{Intersection with Associated Data}).  We sum the total number of TP for collaborators selected by the \emph{Intersection-Size} metric. 
Remember that with the Global Upper Bound (GUB), every victim shares with every other victim and makes perfect predictions about known attackers, i.e., they have access to the ground truth. 
With the Local Upper Bound (LUB), organizations do not share anything but still make perfect predictions based on their local information. 
The accuracy of \emph{Intersection-Size} predictions tends to match LUB, showing that collaboration helps perform as well as a local ``perfect'' predictor, even when considering only $50$ collaboration pairs.

In Table~\ref{tab:improvements}, we summarize the prediction improvement given different benefit estimation metrics, reporting the mean, max, and min improvement, as well as the number of collaborators and coalition size. 
\emph{Pearson} and \emph{Cosine} provide a less significant prediction improvement than set-based metrics. Mean $I$ for \emph{Pearson} and \emph{Cosine} is almost $0.4$, i.e., a (40\% improvement over the baseline), while
mean $I$ for \emph{Jaccard} is close to $0.6$. 
Notably, \emph{Intersection-Size} yields $I$ equal to $1.05$, resulting into a 105\% improvement over the baseline. 
Naturally, the improvement can also be measured for each entity: Maximum improvement with \emph{Intersection-Size} is as high as 700\%.

\begin{figure}[t!]
\centering
\includegraphics[trim=0 0 0 0, scale=0.32]{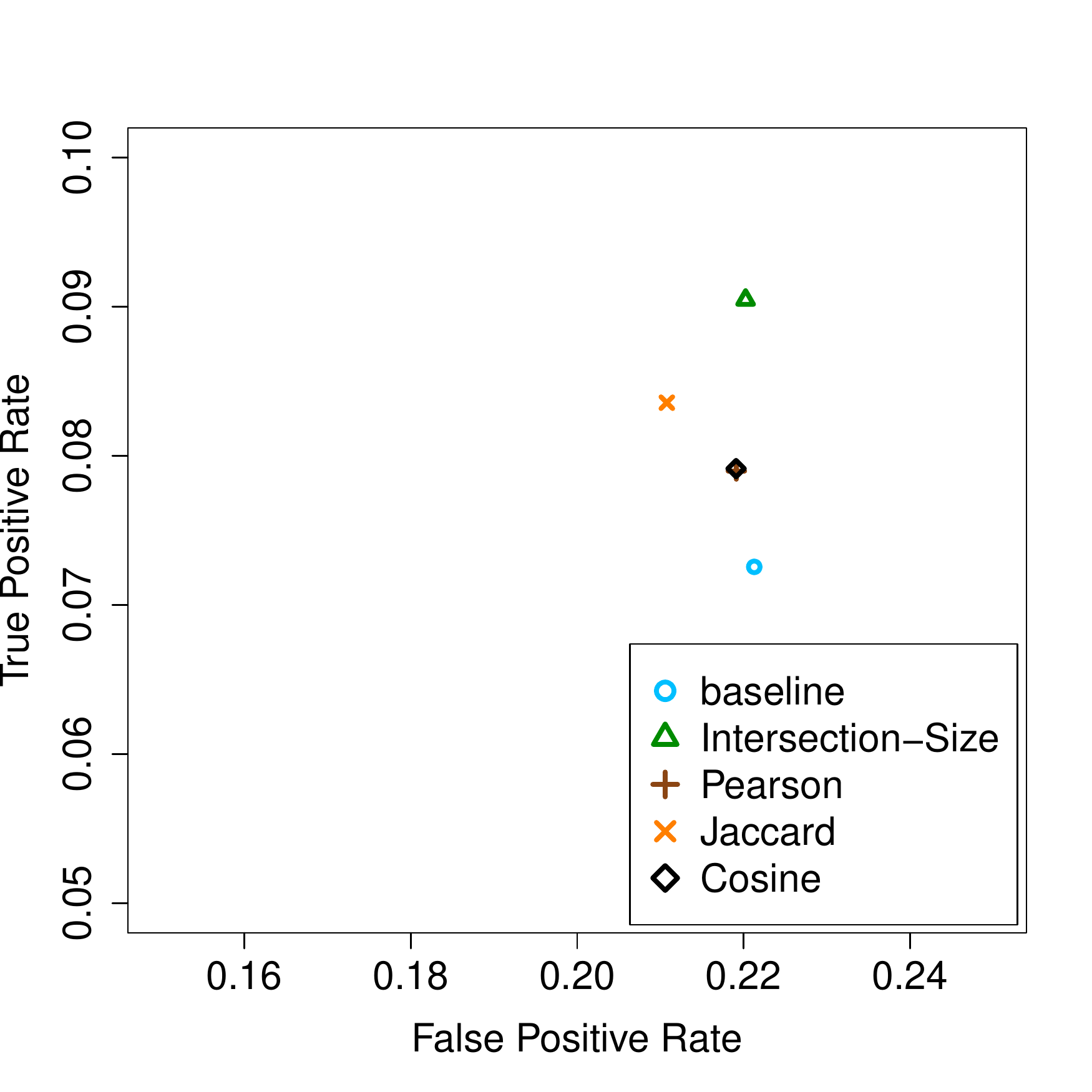}
\vspace{-0.15cm}
\caption{ROC. The x-axis shows the FP rate defined as $FPR = FP / (FP + TN)$, where TN is the number of true negatives. The y-axis shows TP rate which is defined as $TPR = TP / (TP + FN)$, where FN is the number of false negatives.}\label{fig:roc}
\vspace{0.1cm}
\end{figure}

\descr{False Positives.} Fig.~\ref{fig:roc} plots a Receiver Operating Characteristic (ROC) with 
the true positive rate (TPR) against the false positive rate (FPR) for different
benefit estimation strategies (using \emph{Intersection with Associated Data} to share data).
Ideally, we would like to obtain values in the top-left corner (i.e., high TPR and low FPR). 
Interestingly, we observe that all sharing methods improve over the baseline (i.e., they are on upper-left of baseline), thus improving the TPR and reducing the FPR. 
This is a positive result indicating that controlled data sharing helps the prediction system perform better. 

When using \emph{Intersection-Size} to estimate benefits of sharing, TPR improves the most, but FPR does not decrease significantly, whereas \emph{Jaccard} significantly reduces FPR at the cost of a lower TPR increase. 
We also measure the average number of FP with \emph{Intersection-Size} and obtain an average increase of 4\% over the baseline.
In case data is shared using \emph{Union with Associated Data}, FPR decreases even more, but at the cost of an average increase in FP of 55\% with \emph{Intersection-Size}.

\begin{table}[t]
\centering
\small
\begin{tabular}{ | c | c | c | c | c | c | c | c | c |}
  \hline                        
  {\bf Benefit Esti-} & \multicolumn{3}{c |}{\textbf{Improvement}} & \multicolumn{2}{c |}{\textbf{\# Collaborators}} & \multicolumn{3}{c |}{\textbf{Coalition Size}} \\
  \cline{2-4}   \cline{5-6} \cline{7-9}
   {\bf mation Metric} &  Mean  & Max & Min  & Mean & SD  & Mean & SD & Median \\
  \hline                        
    \hline                        
  \emph{Int-Size} & 1.05  & 7 & 0 & 19.47 & 2.24 & 5.09 & 4.09 & 4 \\
  \hline
  \emph{Jaccard} & 0.58 & 8 & 0 & 30.17 & 4.44  & 3.16 & 2.74 & 2 \\
  \hline
\emph{Pearson} & 0.37 & 8 & 0 & 18.08 & 1.40  & 5.20 & 3.15 & 5 \\
  \hline    
  \emph{Cosine} & 0.39 & 8 & 0 & 17.98 & 1.29  & 5.26 & 3.14 & 5  \\
  \hline    
\end{tabular}
\vspace{-0.1cm}
\caption{Fraction of prediction improvements $I$ for collaborators, the number of collaborators, and the size of coalitions. SD stands for Standard Deviation. 
}
\label{tab:improvements}
\end{table}

\begin{figure}[t]
\vspace{-0.1cm}
\centering
   \includegraphics[trim=0 0 0 0, scale =0.35] {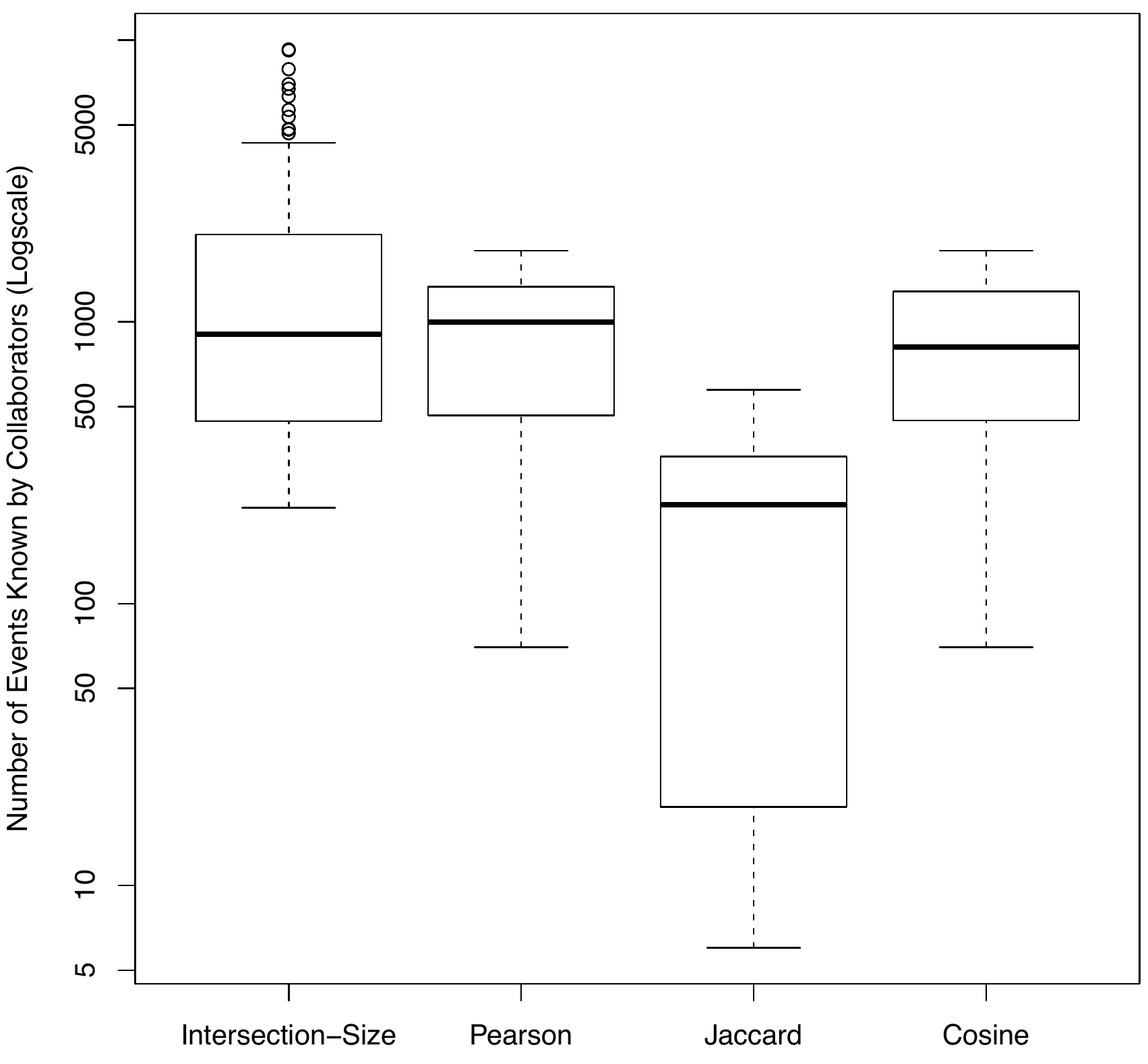}
\vspace{-0.1cm}
\caption{Boxplot of number of events known by collaborators given different benefit estimation metrics. The bottom and top of the box correspond to first and third quartiles. The band inside the box is the second quartile (the median). Outliers are shows with small circles. 
}
   \label{fig:boxplot}
\end{figure}

\subsection{Analysis}

First, we observe that metrics with a normalization factor (i.e., all but \emph{Intersection-Size}) tend to favor partnerships with small collaborators. \emph{Intersection-Size} leads to better performance because it promotes collaboration with larger victims. To confirm this hypothesis, we measure the set size of collaborators according to different metrics (Fig.~\ref{fig:boxplot}) and confirm that metrics with a normalization factor tend to suggest partnerships with collaborators that know less. 
Second, \emph{Pearson} and \emph{Cosine} tend to select partners that are \emph{too} similar: Maximum correlation values are close to $1$, whereas maximum \emph{Jaccard} values only reach $0.5$. Although this implies that targets learn to better defend against specific adversaries, it also leads to little acquired knowledge. 
Third, depending on the metric, entities may partner with previous collaborators, or with new ones. We find that \emph{Intersection-Size}, \emph{Pearson}, and \emph{Cosine} lead to stable groups of collaborators with about 90\% reuse over time, whereas \emph{Jaccard} 
has larger diversity of collaborators over time. This is because about 20\% of victims have high \emph{Jaccard} similarity compared to 4\% for \emph{Pearson} and \emph{Cosine}, providing a larger pool of potential collaborators. 
Hence, if \emph{Intersection-Size} helps a few learn a lot, \emph{Jaccard} helps many victims over time. 

\descr{Statistical Analysis.} A t-test analysis shows that the mean of the number of events known by collaborators differs significantly ($p < 0.0005$) across all pairs of benefit estimation metrics but \emph{Cosine} and \emph{Pearson}. 
If one categorizes collaborators as ``large'' if they have seen more than $500$ events, and ``small'' otherwise, and consider \emph{Cosine} and \emph{Pearson} as one (given the t-test result), we obtain a $3$x$2$ table of benefit estimation metrics and size categories. A $\chi^2$-test shows that categorization differences are statistically significant: \emph{Intersection-Size} tends to select larger collaborators, but also more collaborators than \emph{Pearson/Cosine} (see Table~\ref{tab:improvements}). Other metrics tend to select small collaborators. We obtain $\chi^2(2, N=448)=191.99, p < 0.0005$, where $2$ is the degrees of freedom of the $\chi^2$ estimate, and $N$ is the total number of observations.

\descr{Coalitions.} Recall that, at each time step, different benefit estimation strategies lead to different partnerships in our analysis. Table~\ref{tab:improvements} shows the mean, Standard Deviation (SD), and median number of collaborators per party for different collaboration metrics. We observe that with \emph{Jaccard}, coalitions are smaller and thus entities tend to select less collaborators. 
Other metrics tend to have similar behavior and lead entities to collaborate with about $5$ other entities out of $100$. This is in line with previous work~\cite{katti2005collaborating}, which showed the existence of small groups of correlated entities.  
We also observe that, after a few days (usually $2$), \emph{Intersection-Size}, \emph{Pearson}, and \emph{Cosine} converge to a relatively stable group of collaborators. From one time-step to another, parties continue to collaborate with about 90\% of entities they previously collaborated. In other words, coalitions are relatively stable over time. Comparatively, \emph{Jaccard} has a larger diversity of collaborators over time.

\subsection{Different Sharing Strategies}
\begin{figure}[t]
\centering
\includegraphics[scale=0.32]{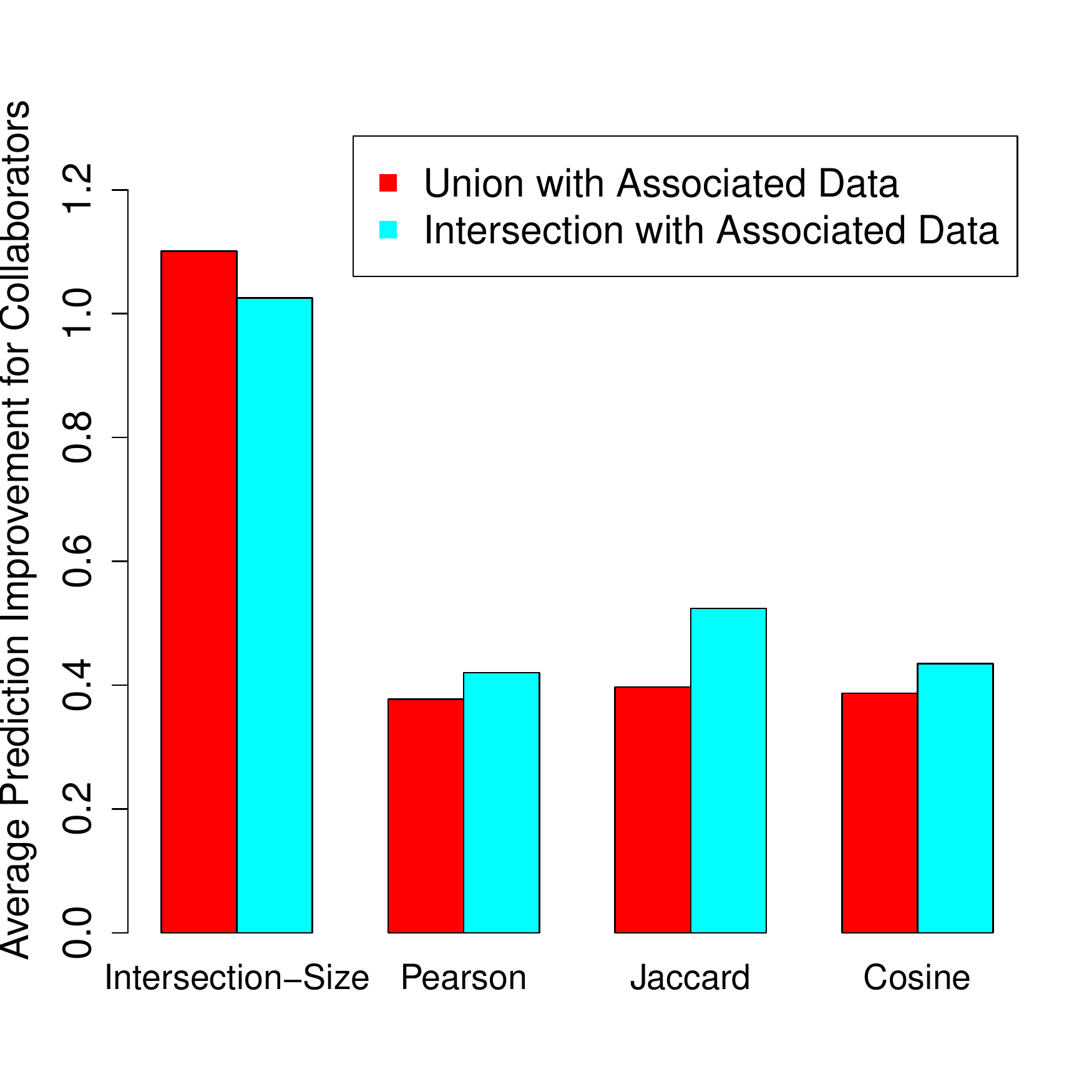}
\vspace{-0.5cm}
\caption{Average prediction improvement for two different data sharing strategies: \emph{Union/Intersection with Associated Data.}}
\label{fig:union}
\end{figure}

The next step is to compare the average prediction improvement $I$ resulting from different data sharing strategies. As showed in Fig.~\ref{fig:union}, \emph{Intersection with Associated Data} performs almost as good as {\em Union with Associated Data} with all benefit estimation metrics. It performs even better when using {\em Jaccard}.

Sharing using the union entails sharing more information, thus, one would expect it to always perform better---however, organizations quickly converge to a stable set of collaborators, and obtain a potentially lower diversity of insights over time. With most metrics, the set of collaborators is stable over time in any case, and so union does perform better than intersection. As previously discussed, \emph{Jaccard} tends to yield a larger diversity of collaborators over time and thus benefits more from \emph{Intersection with Associated Data} as it re-enforces such diversity of insights.

\subsection{Performance of Cryptographic Tools}\label{subsec:perf}
We estimate the operational cost of using cryptographic techniques
for the secure computation of the benefit estimation and the data sharing
routines.

Excluding \emph{Pearson} and \emph{Cosine}  (due to lower accuracy improvement), the protocols for privately estimating benefits of collaborating ({\em Intersection-Size} and {\em Jaccard}) can all be realized based on Private Set Intersection Cardinality (PSI-CA). We chose the instantiation proposed in \cite{DGT12}, which incurs computation and communication overhead linear in sets size. Privacy-preserving data sharing, i.e., 
{\em Intersection with Associated Data}, is instantiated using the PSI-DT protocol from~\cite{DT10}. 
We implemented protocols in C, using GMP and OpenSSL cryptographic libraries, and measured total running times using two Intel Xeon desktops with 3.10GHz CPU connected by a 100Mbps Ethernet link. 
Using sets of size $200$, it takes approximately $400ms$ to execute PSI from~\cite{DT10} and $550ms$ for the PSI-CA from~\cite{DGT12}. Assuming that $100$ organizations are possible partners, there would be $100\cdot99/2$ pairwise executions of PSI-CA and PSI-DT in the worst case, yielding a total running time close to $54s$ for PSI-CA and $40s$ for PSI-DT. That is, it would take less than one minute for one entity to estimate benefits, using PSI-CA, with all other ($99$) parties, and also less than one minute to share data with all possible $99$ partners.
In summary, overhead is appreciably low and could accommodate real-world scenarios where interactions occur several times a day.

\subsection{Take-aways}\label{sec:discussion}
Our experiments confirm that targets that know more tend to successfully predict more attacks. 
However, as indiscriminate sharing poses serious confidentiality, privacy, trust, and liability challenges,
we have considered a controlled data sharing approach aiming to identify partners that help most. 
In our experiments, \emph{Intersection-Size} proves to be the best metric to estimate the benefits of collaborating.
Interestingly, we find that if victims' datasets are very similar, data sharing yields little gain, since there is little to learn. 
This is reinforced by the fact that similarity metrics with a normalization factor favor victims with small datasets.

We find that sharing data with partners using \emph{Intersection with Associated Data} performs almost as good as sharing everything (\emph{Union}).
Not only does intersection provide convenient privacy properties, it also indicates that there is more value in learning about current attackers than other potential attack sources. 
Intuitively, intersection reinforces knowledge about attackers known to a victim, whereas, union might help victims targeted by varying group of attackers. 
Thus, victims benefit as much from improving their knowledge about current attackers, as learning about other sources (that could possibly attack them next). 
In brief, good partners are related but not identical, and should share information about known past attackers.

\descr{Limitations.}
The DShield dataset used in our experiments might be biased toward small organizations that {\em voluntarily} report attack data.
Thus, it might not be directly evident how to generalize our results.
However, our findings indicate that controlled data sharing can remarkably improve prediction, and show statistical evidence that different collaboration strategies affect performance in interesting ways. 
We also make a few simplifying assumptions in our experimental setup, e.g., sampling $100$ random organizations from the Dshield dataset, and establish partnerships by selecting the top 1\% pairs in the benefit estimation matrix. Although we leave the evaluation of the different partnership strategies 
as part of future work, our choices are conservative, thus yielding lower-bound estimates of the benefits of collaboration.

\section{Conclusion}\label{sec:conclusion}
We investigated the viability of a controlled data sharing approach
to collaborative threat mitigation. We focused on collaborative predictive blacklisting
and explored how organizations could quantify expected benefits in a privacy-preserving way (i.e., without disclosing their datasets) before deciding whether or not to share data, and how much.
We performed an empirical evaluation on a dataset of 2 billion suspicious IP address, contributed
by $188$,$522$ organizations to DShield.org over a period of two months. We observed a significant improvement in prediction accuracy (up to 105\%, even when only 1\% of all possible partners collaborate), along with a reduction in the false positive rate.

Our analysis showed that some collaboration strategies work better than others. The number of common attacks provides a good estimation of the benefits of sharing, as it drives entities to partner with more knowledgable collaborators. In fact, only sharing information about common attacks proves to be almost as useful as sharing everything. Our work is the first to show that collaborative threat mitigation does not have to be an ``all-or-nothing'' process: By relying on efficient cryptographic protocols, organizations can share only relevant data, and only when beneficial. 

As part of future work, we intend to study other metrics for benefit estimation (e.g., dissimilarity, data quality~\cite{freudiger2014privacy}) and experiment with other prediction algorithms. We also plan to study and experiment with distributed partner selection strategies, possibly relying on the stable roommate matching problem~\cite{gusfield1989stable}. Finally, we will explore how to adapt our approach to other collaborative security problems, e.g., spam filtering~\cite{damiani2004p2p}, malware detection~\cite{hailpern2001collaborative}, or DDoS mitigation~\cite{oikonomou2006framework}.

\smallskip\noindent{\bf Acknowledgments.} We wish to thank DShield.org and Johannes Ullrich for providing the dataset used in our experiments, as well as Ersin Uzun, Marshall Bern, Craig Saunders, and Anton Chuvakin for their useful comments and feedback. Work done in part while Emiliano De Cristofaro was with PARC.

\appendix \label{app:dshield}
\section{Additional Analysis of the DShield Dataset}

\begin{figure*}[bb]
\centering
\subfigure[]{
   \includegraphics[trim=0 0 0 0, scale =0.32] {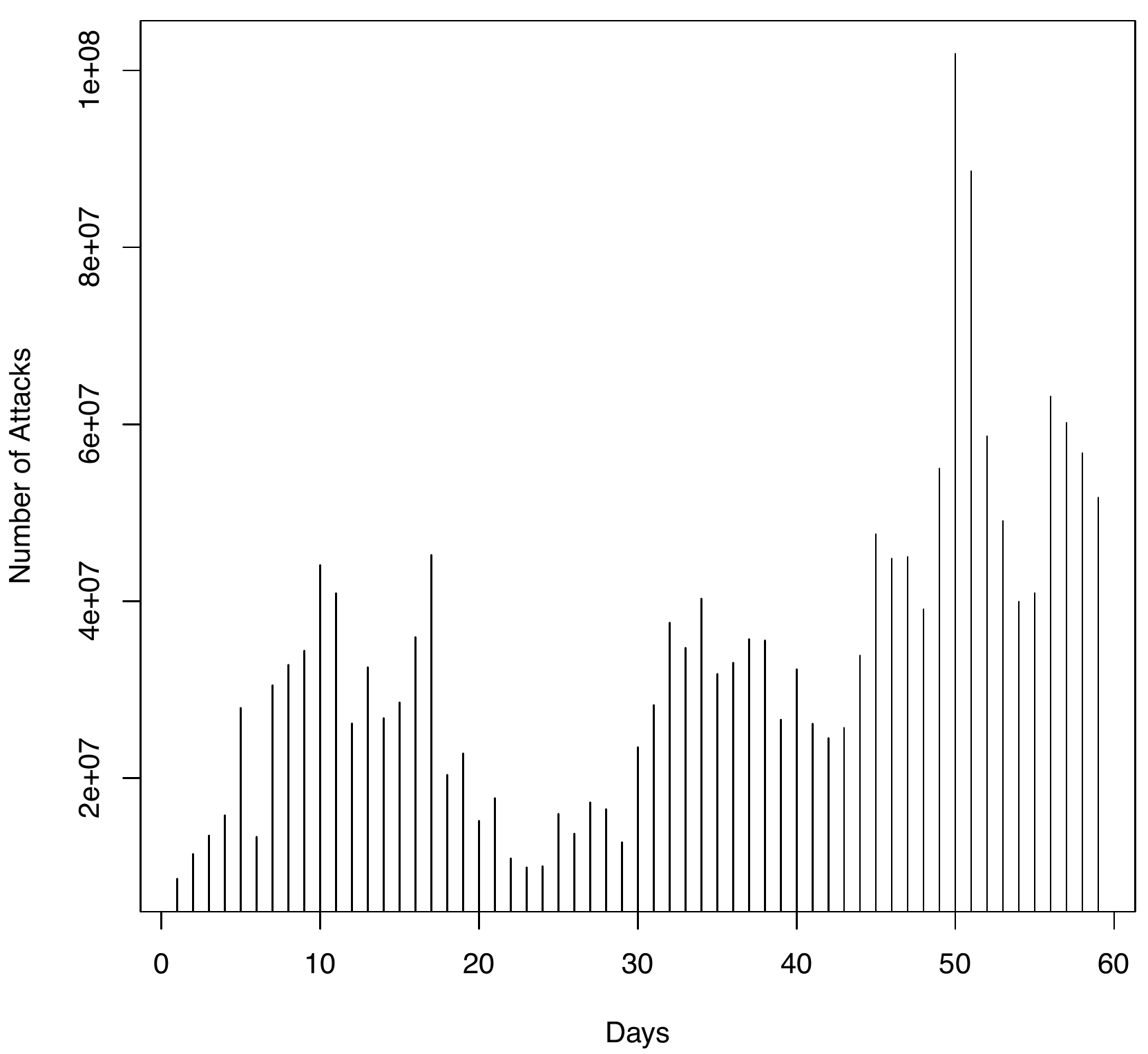}
   \label{fig:general:subfig1}
 }
 \hspace{0.5cm}
  \subfigure[]{
   \includegraphics[trim=0 0 0 0, scale =0.32] {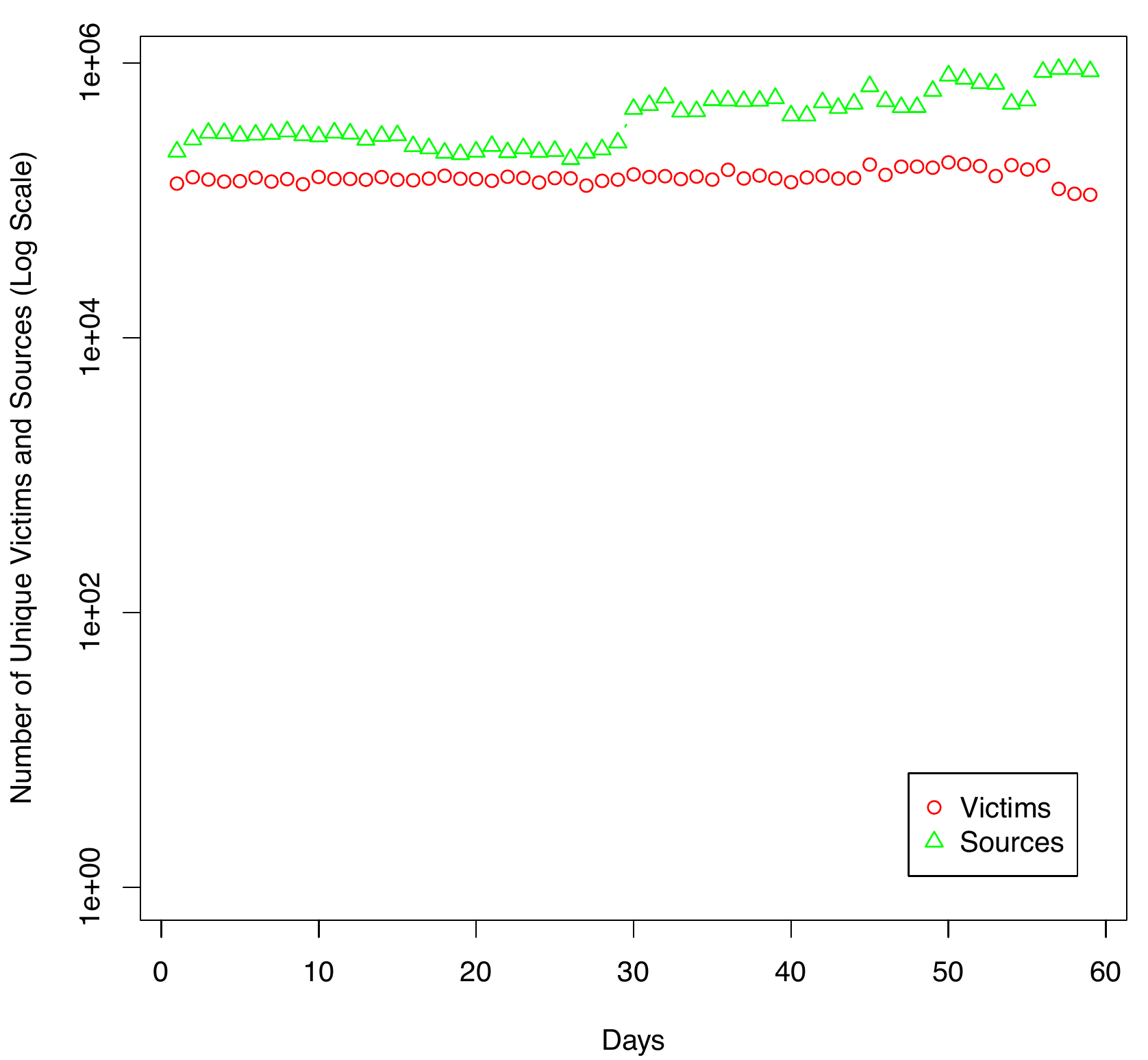}
   \label{fig:general:subfig3}
 }
  \vspace{-0.3cm}
\caption{General DShield characteristics: (a) Histogram of number of attacks per day. (b) Number of unique targets and sources.}
\end{figure*}

\descr{General Statistics.}
We present in Fig.~\ref{fig:general:subfig1} the histogram of the number of attacks per day, indicating about $30$M daily attacks. We observe a significant increase around day $50$ to $100$M attacks. Careful analysis reveals that a series of IP addresses starts to aggressively attack around day $50$, indicating a possible DoS attack initiation.  

Fig.~\ref{fig:general:subfig3} shows the number of unique targets and sources over time. Detailed analysis shows a stable number of sources and targets. This stability confirms that it should be possible to predict attackers' tactics based on past observations.
An analysis of attacked ports shows that top 10 attacked ports (with more than 10M hits) are Telnet, HTTP, SSH, DNS, FTP, BGP, Active Directory, and Netbios ports. This shows a clear trend towards misuse of popular web services.

\begin{figure*}[ttt]
	\centering
\begin{minipage}[t]{0.45\linewidth}
	\centering
   \includegraphics[trim=0 0 0 0, scale =0.32] {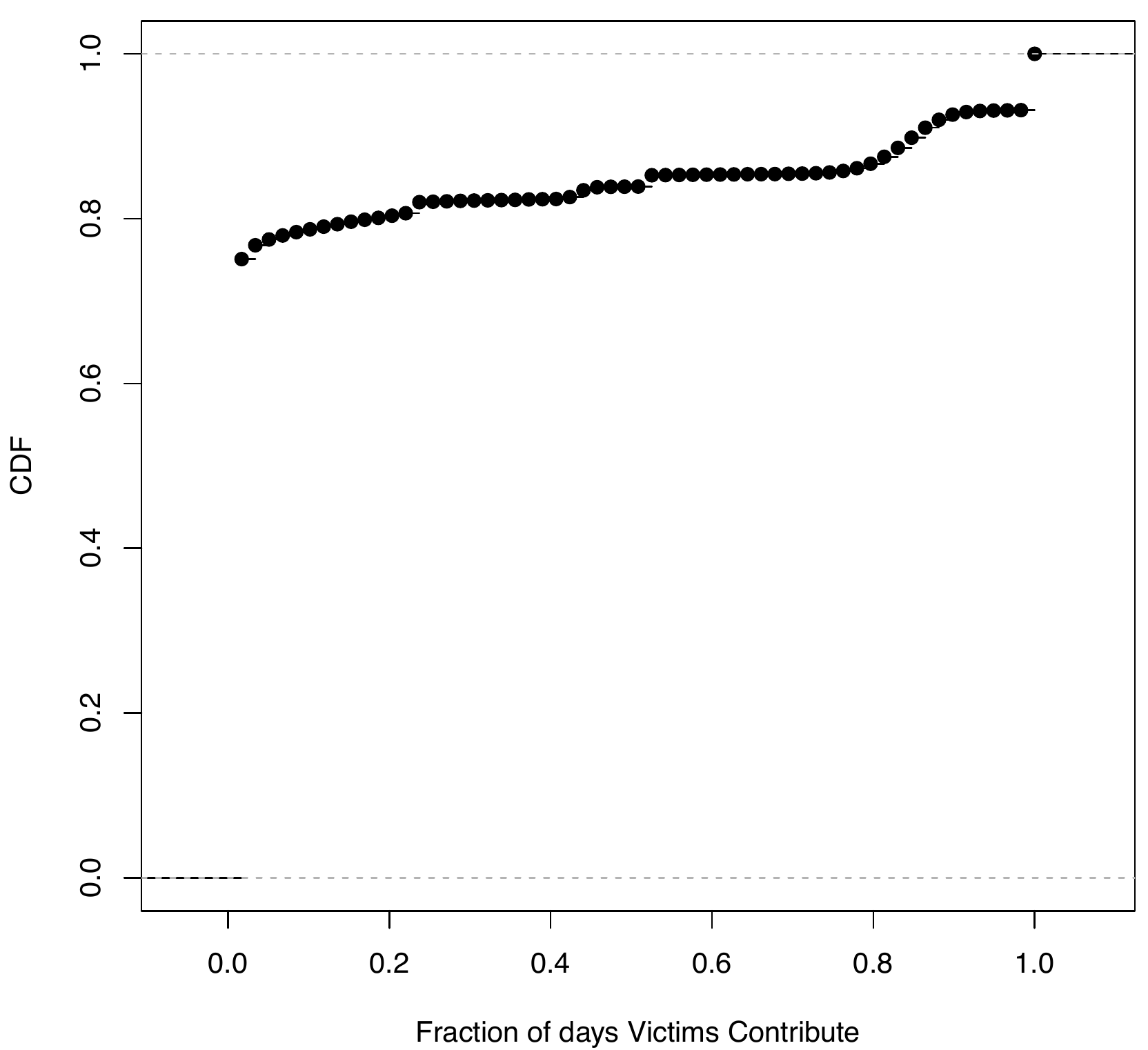}
   \vspace{-0.2cm}
	\caption{Fraction of days each target contributes. }
	\label{fig:general}
\end{minipage}
\hspace{0.3cm}
\begin{minipage}[t]{0.45\linewidth}
   \includegraphics[trim=0 0 0 0, scale =0.32] {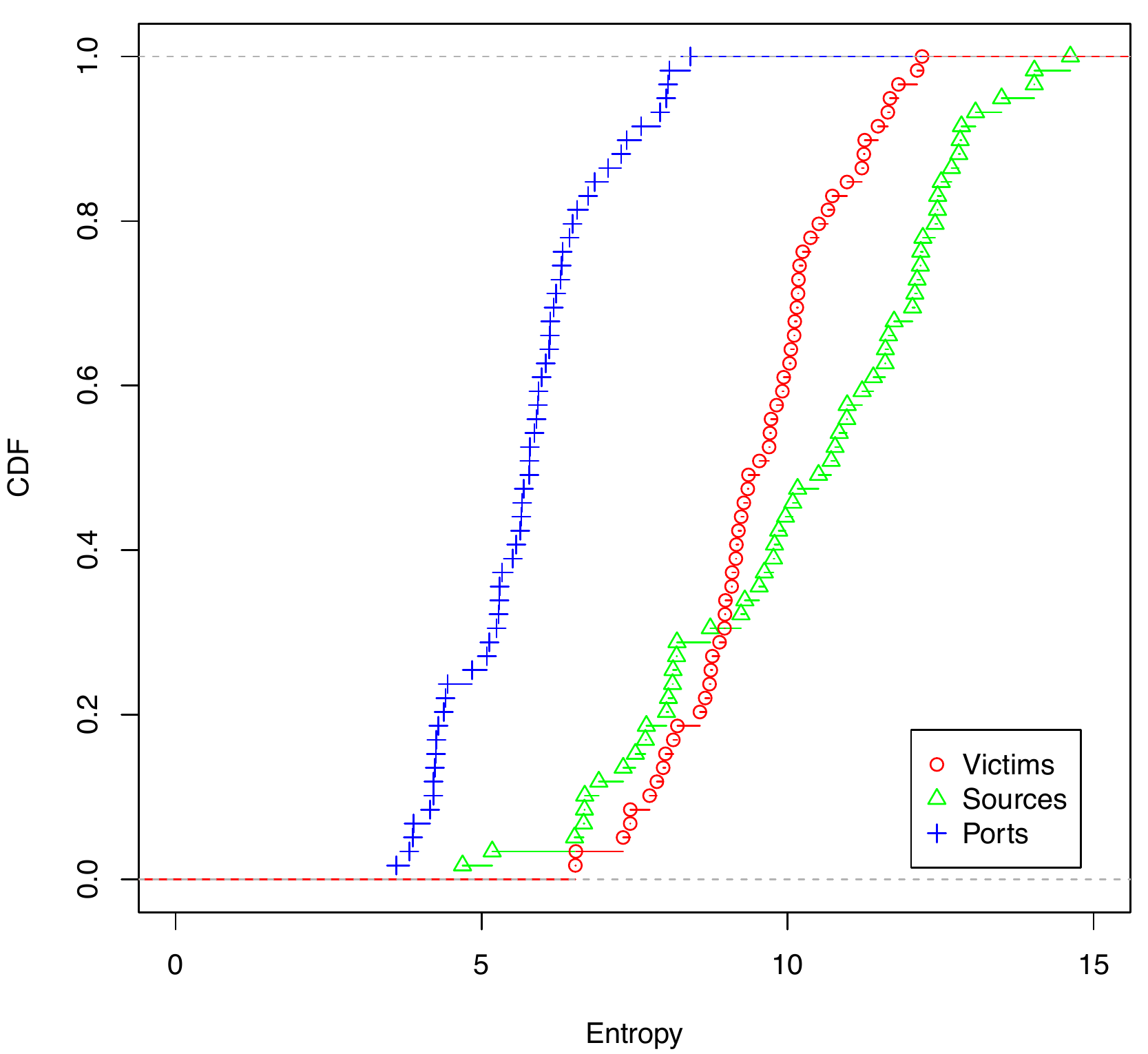}
   \centering
   \vspace{-0.2cm}
   \caption{CDF of entropy of different attack parameters.}
      \label{fig:entropy}
\end{minipage}
 \end{figure*}

In Fig.~\ref{fig:general}, we plot the CDF of the fraction of victims that contribute logs to DShield over the course of two months, and observe that few victims contribute daily.

\descr{Predictability.} Fig.~\ref{fig:entropy} shows the CDF of the Shannon entropy of the different log entry elements. It helps us visualize the uncertainty about a given IP address, port number or target appearing in the logs, and thus estimate our ability to predict those values. To obtain this figure, we estimate the probability of each victim, source or port being attacked each day. For example, for each port $i$, we compute: 
\begin{equation}
\small
\mbox{Pr}(\mbox{Port } i \mbox{ on day } j ) = \frac{\mbox{Attacks on Port } i \mbox{ on day } j} {\mbox{Attacks on day } j} 
\end{equation}   

\noindent We also compute the entropy for each day and aggregate it overall using the CDF. 
Previous work~\cite{song2010limits} showed that, following Fano's inequality,  entropy correlates with predictability. We observe that ports numbers have the lower entropy distribution, indicating a small set of targeted ports: $80\%$ of attacks target a set of $2^7=128$ ports, indicating high predictability. We also observe that victims are more predictable than sources, as $90\%$ of victims lie within a set of $2^{12}=4096$ victims as compared to $90\%$ of sources being in a list of $2^{14}=16,384$ sources. Victims' set is thus significantly smaller and more predictable than attackers' set. 

\begin{figure*}[hhh]
\centering
\centering
\subfigure[]{
   \includegraphics[trim=0 0 0 0, scale =0.29] {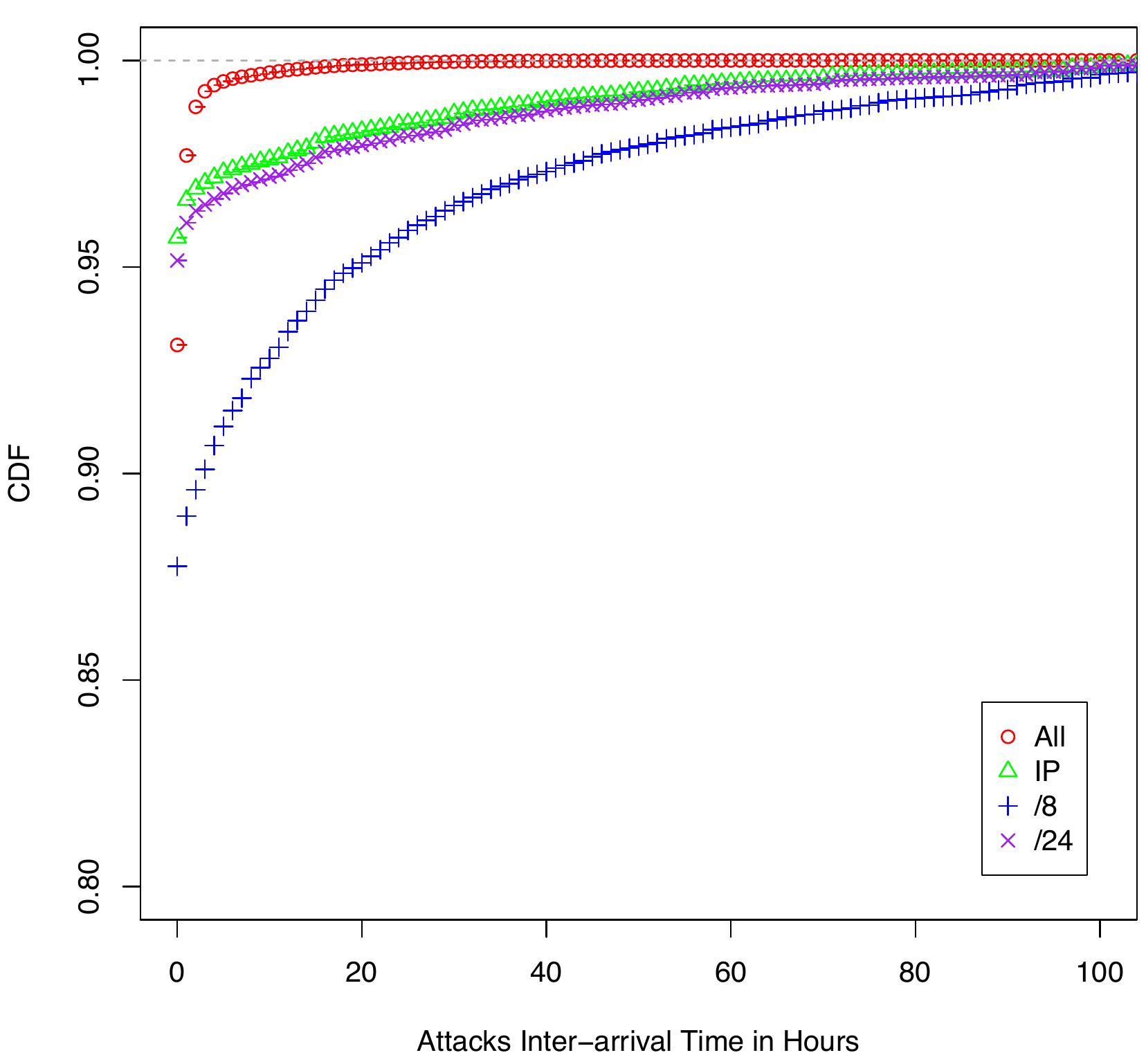}
   \label{fig:interarrival:subfig1}
 }
   \hspace{0.2cm}
 \subfigure[]{
   \includegraphics[trim=0 0 0 0, scale =0.29] {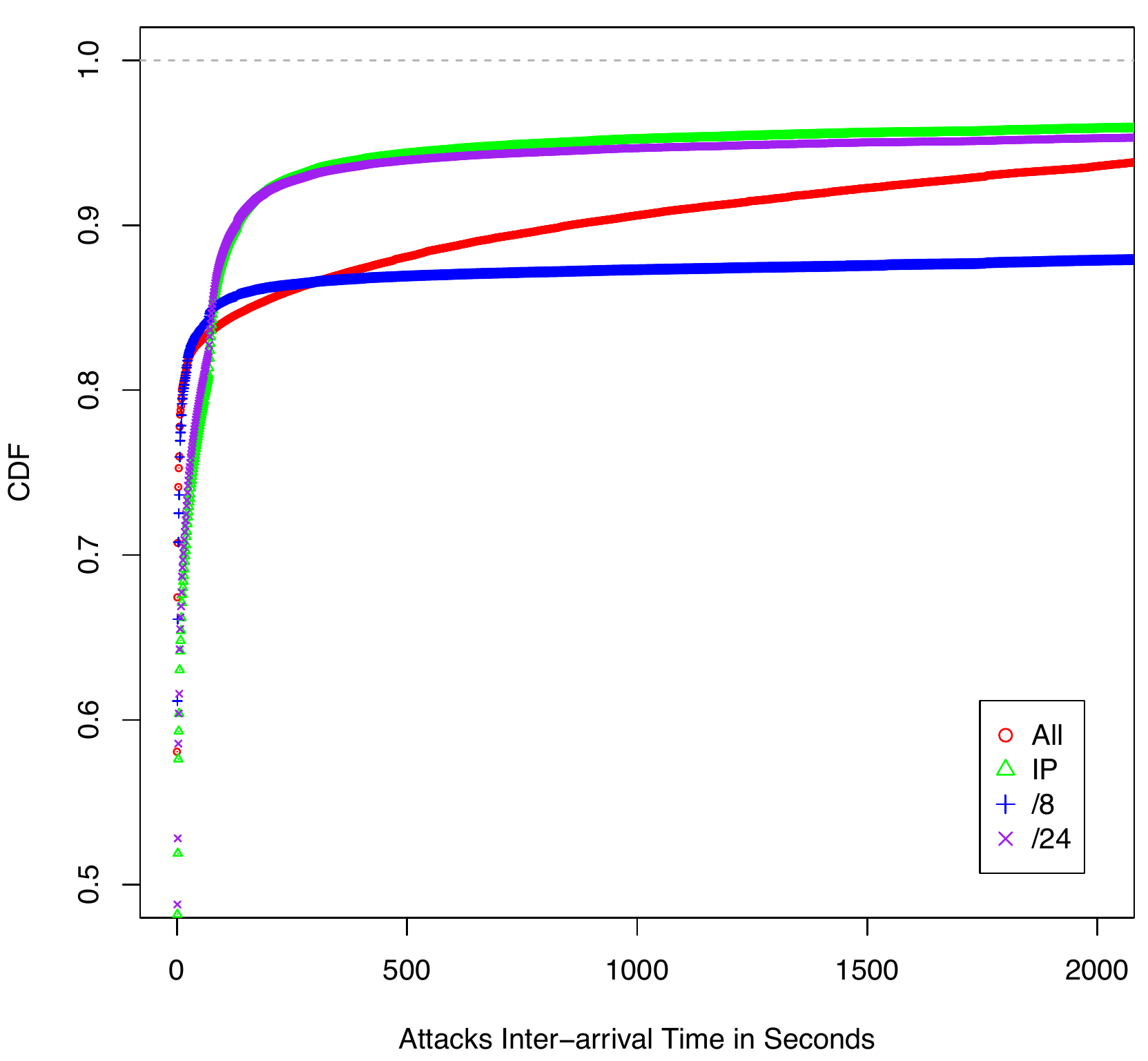}
   \label{fig:interarrival:subfig2}
 }
\label{fig:interarrival}
\caption{CDF of inter-arrival time of attacks: (a) Per hour, and (b) per second. All indicates the inter-arrival time of any attacks, $/8$ of common $/8$ subnetworks, $/24$ of common $/24$ subnetworks,  and IP of the same IP. }
\end{figure*}

\descr{Intensity.}
 Fig.~\ref{fig:interarrival:subfig1} shows the inter-arrival time of attacks in hours, and Fig.~\ref{fig:interarrival:subfig2} shows the inter-arrival time of attacks in seconds. We observe that almost all attacks occur within 3-minute windows. IP addresses and $/24$ subnetworks have similar behavior. In particular, Fig.~\ref{fig:interarrival:subfig2} shows that in short time intervals, $85\%$ of $/8$ subnetworks have short attack inter-arrival time indicating the bursty attacks on such networks. Attackers target subnetworks for a short time and then disappear.

\vspace{-0.1cm}


\begin{thebibliography}{10}

\bibitem{fb}
{Facebook ThreatExchange}.
\newblock \url{https://threatexchange.fb.com}, 2015.

\bibitem{guardian}
S.~Ackerman.
\newblock {Privacy experts question Obama's plan for new agency to counter
  cyber threats -- The Guardian}.
\newblock \url{http://gu.com/p/45yvz}, 2015.

\bibitem{adar2007user}
E.~Adar.
\newblock {User 4xxxxx9: Anonymizing query logs}.
\newblock In {\em Query Log Analysis Workshop}, 2007.

\bibitem{pets10}
B.~Applebaum, H.~Ringberg, M.~Freedman, M.~Caesar, and J.~Rexford.
\newblock {Collaborative, privacy-preserving data aggregation at scale}.
\newblock In {\em PETS}, 2010.

\bibitem{bilogrevic2014s}
I.~Bilogrevic, J.~Freudiger, E.~De~Cristofaro, and E.~Uzun.
\newblock {What's the gist? Privacy-preserving aggregation of user profiles}.
\newblock In {\em ESORICS}, 2014.

\bibitem{espresso}
C.~Blundo, E.~De~Cristofaro, and P.~Gasti.
\newblock {EsPRESSo: Efficient privacy-preserving evaluation of sample set
  similarity}.
\newblock {\em JCS}, 22(3), 2014.

\bibitem{burkhart2010sepia}
M.~Burkhart, M.~Strasser, D.~Many, and X.~Dimitropoulos.
\newblock {SEPIA: Privacy-preserving aggregation of multi-domain network events
  and statistics}.
\newblock In {\em Usenix Security}, 2010.

\bibitem{coull2007playing}
S.~E. Coull, C.~V. Wright, F.~Monrose, M.~P. Collins, and M.~K. Reiter.
\newblock {Playing Devil's Advocate: Inferring sensitive information from
  anonymized network traces}.
\newblock In {\em NDSS}, 2007.

\bibitem{CSRIC}
{CSRIC Working Group 7}.
\newblock {U.S. anti-bot code of conduct for Internet service providers:
  Barriers and metrics considerations}, 2013.

\bibitem{damiani2004p2p}
E.~Damiani, S.~De~Capitani~di Vimercati, S.~Paraboschi, and P.~Samarati.
\newblock {P2P-based collaborative spam detection and filtering}.
\newblock In {\em P2P}, 2004.

\bibitem{DGT12}
E.~De~Cristofaro, P.~Gasti, and G.~Tsudik.
\newblock {Fast and private computation of cardinality of set intersection and
  union}.
\newblock In {\em CANS}, 2012.

\bibitem{DT10}
E.~{De Cristofaro} and G.~Tsudik.
\newblock Practical private set intersection protocols with linear complexity.
\newblock In {\em FC}, 2010.

\bibitem{DT12}
E.~{De Cristofaro} and G.~Tsudik.
\newblock Experimenting with fast private set intersection.
\newblock In {\em TRUST}, 2012.

\bibitem{FNP}
M.~Freedman, K.~Nissim, and B.~Pinkas.
\newblock Efficient private matching and set intersection.
\newblock In {\em EUROCRYPT}, 2004.

\bibitem{freudiger2014privacy}
J.~Freudiger, S.~Rane, A.~E. Brito, and E.~Uzun.
\newblock {Privacy Preserving Data Quality Assessment for High-Fidelity Data
  Sharing}.
\newblock In {\em WISCS}, 2014.

\bibitem{gusfield1989stable}
D.~Gusfield and R.~W. Irving.
\newblock {\em The stable marriage problem: structure and algorithms}.
\newblock MIT Press Cambridge, 1989.

\bibitem{hailpern2001collaborative}
B.~T. Hailpern, P.~K. Malkin, and R.~J. Schloss.
\newblock Collaborative server processing of content and meta-information with
  application to virus checking in a server network, 2001.
\newblock US Patent 6,275,937.

\bibitem{HEK12}
Y.~Huang, D.~Evans, and J.~Katz.
\newblock {Private Set Intersection: Are Garbled Circuits better than custom
  protocols?}
\newblock In {\em NDSS}, 2012.

\bibitem{HEKM11}
Y.~Huang, D.~Evans, J.~Katz, and L.~Malka.
\newblock {Faster secure two-party computation using Garbled Circuits}.
\newblock In {\em Usenix Security}, 2011.

\bibitem{jaccard}
P.~Jaccard.
\newblock {Etude comparative de la distribution florale dans une portion des
  Alpes et du Jura}.

\bibitem{katti2005collaborating}
S.~Katti, B.~Krishnamurthy, and D.~Katabi.
\newblock Collaborating against common enemies.
\newblock In {\em IMC}, 2005.

\bibitem{kenneally2010dialing}
E.~Kenneally and K.~Claffy.
\newblock {Dialing privacy and utility: A proposed data-sharing framework to
  advance Internet research}.
\newblock {\em IEEE Security \& Privacy}, 8(4), 2010.

\bibitem{lakkaraju2008evaluating}
K.~Lakkaraju and A.~Slagell.
\newblock Evaluating the utility of anonymized network traces for intrusion
  detection.
\newblock In {\em Securecomm}, 2008.

\bibitem{lincoln2004privacy}
P.~Lincoln, P.~Porras, and V.~Shmatikov.
\newblock Privacy-preserving sharing and correction of security alerts.
\newblock In {\em Usenix Security}, 2004.

\bibitem{locasto2005towards}
M.~E. Locasto, J.~J. Parekh, A.~D. Keromytis, and S.~J. Stolfo.
\newblock {Towards collaborative security and P2P intrusion detection}.
\newblock In {\em Information Assurance Workshop}, 2005.

\bibitem{oikonomou2006framework}
G.~Oikonomou, J.~Mirkovic, P.~Reiher, and M.~Robinson.
\newblock {A framework for a collaborative DDoS defense}.
\newblock In {\em ACSAC}, 2006.

\bibitem{PSZ14}
B.~Pinkas, T.~Schneider, and M.~Zohner.
\newblock Faster private set intersection based on {OT} extension.
\newblock In {\em Usenix Security}, 2014.

\bibitem{porras2006large}
P.~Porras and V.~Shmatikov.
\newblock Large-scale collection and sanitization of network security data:
  risks and challenges.
\newblock In {\em New Security Paradigms Workshop (NSPW)}, 2006.

\bibitem{pouget2005vh}
F.~Pouget, M.~Dacier, and V.~H. Pham.
\newblock Vh: Leurre. com: on the advantages of deploying a large scale
  distributed honeypot platform.
\newblock In {\em E-Crime and Computer Conference}, 2005.

\bibitem{redsky}
{Red Sky Alliance}.
\newblock \url{http://redskyalliance.org/}.

\bibitem{dshield}
{SANS Technology Institute}.
\newblock {DShield Data}.
\newblock \url{https://www.dshield.org/}.

\bibitem{slagell2005sharing}
A.~Slagell and W.~Yurcik.
\newblock Sharing computer network logs for security and privacy: A motivation
  for new methodologies of anonymization.
\newblock In {\em Securecomm}, 2005.

\bibitem{soldo}
F.~Soldo, A.~Le, and A.~Markopoulou.
\newblock Predictive blacklisting as an implicit recommendation system.
\newblock In {\em INFOCOM}, 2010.

\bibitem{song2010limits}
C.~Song, Z.~Qu, N.~Blumm, and A.-L. Barab{\'a}si.
\newblock Limits of predictability in human mobility.
\newblock {\em Science}, pages 1018--1021, 2010.

\bibitem{whitehouse_last}
{The White House}.
\newblock {Executive order promoting private sector cybersecurity information
  sharing}.
\newblock \url{http://1.usa.gov/1vISfBO}, 2015.

\bibitem{wombat}
{Worldwide Observatory of Malicious Behaviors and Attack Threats}.
\newblock \url{http://www.wombat-project.eu/}, 2013.

\bibitem{xu2002prefix}
J.~Xu, J.~Fan, M.~H. Ammar, and S.~B. Moon.
\newblock Prefix-preserving {IP} address anonymization: Measurement-based
  security evaluation and a new cryptography-based scheme.
\newblock In {\em ICNP}, 2002.

\bibitem{Yao}
A.~Yao.
\newblock {Protocols for secure computations}.
\newblock In {\em FOCS}, 1982.

\bibitem{DOMINO}
V.~Yegneswaran, P.~Barford, and S.~Jha.
\newblock {Global intrusion detection in the DOMINO overlay system}.
\newblock In {\em NDSS}, 2004.

\bibitem{zhang2008highly}
J.~Zhang, P.~A. Porras, and J.~Ullrich.
\newblock Highly predictive blacklisting.
\newblock In {\em Usenix Security}, 2008.

\end{thebibliography}
\end{document}